\documentclass[aps]{revtex4}
\usepackage[dvips]{color,graphicx}
\newcommand{\bfx}{{\bf x}}
\newcommand{\bfy}{{\bf y}}
\newcommand{\eq}[1]{Eq.~(\ref{#1})}

\def\bfq {{\bf q}}

\def\bfK{{\bf K}}

\def\bfp{{\bf p}}  
\def\bfu{{\bf u}}
 
\def\bfy{{\bf y}} 
\def\bfx{{\bf x}}  
\def\be{\begin{equation}}
 \def \ee{\end{equation}}
\def\bea{\begin{eqnarray}}
  \def\eea{\end{eqnarray}}
\def\non{\nonumber\\}

\def\bfb{{\bf b}}
\begin{document}
\hfil {NT@UW-05-07 paper5.tex in /hbt/hbtpaper/prcsub}
\title{Polishing the Lens: I Pionic Final State Interactions and  HBT Correlations-- Distorted Wave Emission Function (DWEF) Formalism and Examples}

\author{Gerald A. Miller and John G. Cramer \\
Department of Physics, University of Washington\\
  Seattle, WA 98195-1560}

\begin{abstract}
The emission of pions produced within a dense, strongly-interacting system of matter in the presence of strong 
radial
flow and absorption is described using a relativistic optical model formalism, replacing the attenuated or 
unattenuated
plane waves of earlier emission function approaches with ``distorted wave'' solutions to a relativistic wave 
equation
including a complex optical potential.  The resulting distorted-wave emission function model (DWEF) is used in
numerical calculations to fit HBT correlations and the resonance-corrected pion spectrum from central-collision STAR
Au+Au pion data at $\sqrt{s}= 200$ GeV.
The parameters of the emission function are  constrained by adopting a pion formation temperature taken from
lattice gauge calculations and a  chemical potential equal to the pion mass 
suggested by chiral symmetry restoration.
Excellent agreement with the STAR data are  obtained.  The applications are  extended by applying linear
participant scaling to the space-time parameters of the model while keeping all other parameters
fixed. This allows us to predict
 HBT radii over a range of centralities for both Au+Au and Cu+Cu collisions.  Good agreement
is found with STAR HBT data for all but the most peripheral collisions.
  The squares of  pionic distorted wave functions, obtained as exact
numerical solutions to the wave equation, are displayed and significant differences
with  the results of using the familiar eikonal approximation are found. Using
the eikonal approximation  leads to a  qualitative accounting for the effects of the 
imaginary part of the optical potential, but fails entirely to include the effects of
the real part of the optical potential. A simple example is  used to illustrate
that an attractive optical potential can have large effects on extracting radii and
can also lead to oscillations in  radii measured at low momenta. 
 \end{abstract}

\maketitle
\vskip 0.5in

\section{Introduction}

Measurements of the two-particle momentum correlations
between pairs of identical particles have been used to study
the space-time structure of the 
``fireball'' produced
in the collision between two heavy ions moving relativistically. 
 The quantum statistical 
effects of symmetrization cause an enhancement of the 
two-boson coincidence rate at small momentum differences
that can be related to the space-time extent  of the particle source.
This method, called HBT interferometry,  has been applied extensively 
in recent experiments at the Relativistic
 Heavy Ion Collider (RHIC) by the
STAR and PHENIX 
collaborations. See the reviews\cite{Pratt:wm,Wiedemann:1999qn,Kolb:2003dz,Lisa:2005dd}.

The invariant ratio of the 
 cross section for the production of two 
  pions of momenta $\bfp_1,\bfp_2$
to the product of single particle production cross sections is analyzed as
the correlation function $C(\bfp_1,\bfp_2)$.
 We define
 $\bfq$=$\bfp_1$--$\bfp_2$
and $\bfK$=$(\bfp_1$+$\bfp_2)/2$, with $\bfK_T$ as 
the component perpendicular to the beam direction. 
(We focus on mid-rapidity data, where 
$\bfK=\bfK_T$.)
The correlation function
can be   parameterized
for small $\bfq$ as 
$
C(\bfq,\bfK)-1 \approx \lambda \exp{(-R_O^2q_O^2-R_S^2q_S^2-R_L^2q_L^2)}\approx
\lambda(1-R_O^2q_O^2-R_S^2q_S^2-R_L^2q_L^2)\; (q_iR_i\ll1), $   
where $O,S,L$ 
represent directions parallel to $\bfK_T$, perpendicular
to $\bfK_T$ and the beam direction, 
and parallel to the beam direction\cite{BP-HBT}. 
Early \cite{Rischke:1996em} and recent \cite{Kolb:2003dz} hydrodynamic 
calculations predicted  that  a fireball evolving through a quark-gluon-hadronic phase 
transitions would emit pions
over a long time period, causing a large ratio $R_O/R_S$.
The puzzling experimental
result that $R_O/R_S\approx 1$ 
\cite{Adler:2001zd} is
part of what has been  
 called ``the RHIC HBT puzzle'' \cite{Heinz:2002un}. 
 
Another part of the puzzle is that the measured radii depend strongly
on the average momentum $K$, typically decreasing in size by about 50\% over
the measured range. This dependence of a geometrical parameter on the
probe momentum shows immediately that the radii are not simply a property of a static 
source.
The influence 
of the interactions between the pionic probe and the medium, as well as the effects of transverse flow,
must be taken into account when extracting the radii. 
The medium at RHIC seems 
to be a very high density, strongly interacting plasma\cite{Gyulassy:2004zy},
so that any pions made in its interior would be expected to interact strongly.

We  studied the effects of  including the pionic interactions  
in previous work\cite{Cramer:2004ih}, finding that 
the only way to simultaneously describe the measured 
HBT radii and pionic spectra is to 
include the effects of pion-medium final state interactions by solving the relevant
relativistic wave equation.  These interactions are
so strongly attractive that the pions can be taken as propagating through 
 a system with a restored chiral symmetry.

The principal aim of the present work is to present a detailed treatment of the
formalism  that will allow wide application of our technique, and we present some new
applications here.
 We also 
provide specific simple examples to  demonstrate that the effects of
 pionic interactions cause the measured sizes of the medium to be different 
than the true sizes. Furthermore, we shall explicitly demonstrate that  classical
treatments of the pion-medium interactions, based on using the eikonal approximation for solutions of the wave equation
 are not valid.

An outline of the remainder of this paper follows.
Previous  standard formalisms that use plane wave pions are briefly 
reviewed in Sect.~\ref{sec:pwp}.  The technical method of  incorporating the
influence of final state interactions between the pion and the medium
is described in   Sect.~\ref{sec:fsi}. 
Pionic emission in the absence of final state interaction is  
described with an emission  function 
$S_0$ that is of a form motivated by hydrodynamics. The connection between the
symmetries of this
function (for the  case of head-on collisions) 
and the form of the pion optical potential is described in Sect.~\ref{sec:sss}.
The role of the complex optical potential
and the use  of chiral symmetry 
to constrain its  form at  low energies
 is explained in Sect.~\ref{sec:fsiop}.
The the specific numerical algorithm necessary to  incorporate the optical potential 
is presented  in  Sect.~\ref{sec:psi}. Once our approach is defined, there is 
 only one approximation we need to make. The validity of this requires only that  the
source size  be much larger than the inverse of the temperature.
This large source approximation, LSA,  is explained in Sect.~\ref{sec:dwef}. The 
resulting distorted wave emission function, DWEF, is evaluated
using two different equivalent methods   in Sec.~\ref{sec:corr}.
In Sect.~\ref{sec:apps} we apply these methods to STAR central Au+Au data
at $\sqrt{s}=200$ GeV.  Instead of treating the temperature of the
system as a free parameter, as was done in \cite{Cramer:2004ih},
we fix its value at the transition temperature $T_c =$ 193 MeV, obtained
in the most recent lattice QCD calculation\cite{Katz:2005br}. 

For conventional calculations of the spectra, chemical equilibrium analyzes yield
lower temperatures $T_{ch}=174$ MeV \cite{Braun-Munzinger:2001ip}. A large difference between
$T_c$ and $T_{ch}$ implies that the hadrons interact after the deconfinement
transition occurs. This notion is entirely consistent with our treatment of 
pionic distortions which has  as its fundamental assumption that 
pions interact in a hot dense medium before escaping to freedom.
We also fix the  pion 
chemical potential at the pion mass ($\mu_{\pi} =$ 139.57 MeV). 
The resulting procedure reduces the number of free parameters by two to a total of nine.
Then we are able to reproduce a total of 32 data points with high accuracy.
Section~\ref{sec:apps} also extends earlier numerical results\cite{Cramer:2004ih} by computing the
dependence of the HBT radii on the centrality in Au+Au collisions and
by making predictions for  Cu+Cu collisions vs. centrality.
The eikonal approximation to solving the wave equation is discussed in 
Sect.~\ref{sec:eik} in which  
the important influence of the opacity and the vanishing
of the effects of the real potential are also described.
 The importance of the real part of the
optical potential  in obtaining oscillating radii at low energies 
is illustrated through two examples in Sect.~\ref{sec:wigg}.
A brief summary is presented in  Sect.~\ref{sec:sum}. Finally, 
a short appendix verifies an approximation to an
integral.

\section {Previous Formalism -- Plane wave  pions}
\label{sec:pwp}
The aim is to include the effects of final state interactions of outgoing
pions. We'll begin with a brief review of the formalism previously 
used to describe
HBT
correlations for situations in which the pions do not interact with the medium.

The 
relevant observables are the 
 covariant single-
and two-particle emission functions defined as appropriately normalized ratios 
of cross sections\cite{GKW79}
\bea 
{\cal P}_1(\bfp)=E{1\over\sigma_\pi}{d\sigma_\pi\over d^3\bfp}\\
{\cal P}_2(\bfp_1,\bfp_2)=E_1E_2{1\over\sigma_\pi}{d\sigma_{\pi\pi}\over d^3\bfp_1\;d^3\bfp_2}\\
\int{d^3p\over E}{\cal P}_1(\bfp)=\langle N\rangle\label{spectra0}\\
\int{d^3p_1\over E_1}{d^3p_2\over E_2}{\cal P}_2(\bfp_1,\bfp_2)=\langle N(N-1)\rangle\\
C(\bfp_1,\bfp_2)=\frac{\langle N \rangle^2}
{\langle N(N-1)\rangle}
\frac{{\cal P}_2(\bfp_1,\bfp_2)}{{\cal P}_1(\bfp_1){\cal P}_1(\bfp_2)}
\to 
\frac{{\cal P}_2(\bfp_1,\bfp_2)}{{\cal P}_1(\bfp_1)
{\cal P}_1(\bfp_2)}.\label{cp1p2}
\eea
The last step of Eq.~(\ref{cp1p2}) is obtained because 
$\langle N^2\rangle\gg\langle N\rangle$ for 
the very high energy collisions 
of  interest here.

 These observables can be expressed in terms of an emission
 function $S(x,K)$ that  is the 
Wigner transform of the density matrix associated with the 
currents that  emit  the pions. 
The literature \cite{Wiedemann:1999qn,GKW79}
presents the single-particle emission function as 
\bea &&
S_0(x,p)=\int 
{d^4y\over2(2\pi)^3}\exp{(-i p\cdot y)}\langle {J}^*(x+y/2) {J}(x-y/2)\rangle\non
&& 
=\int 
{d^4y\over2(2\pi)^3}\langle e^{-ip\cdot(x+y/2)} {J}^*(x+y/2) 
 e^{ip\cdot(x-y/2)}{J}(x-y/2)\rangle, \label{pwa}
\eea
where $p$ is  the four-momentum  of an  on-shell emitted pion
and $J(x)$ is the current that acts as a source of pion fields.
We shall assume that $J(x)$ is not altered by the presence of the produced
pions.
 The brackets indicate that one takes an  
 ensemble average over chaotic sources.  Thus we  assume 
 that the emission process is initially
uncorrelated: 
the pions are emitted from chaotic sources that have
 random phases\cite{ed73,GKW79,Kapusta:2005pt}.  
This assumption seems consistent with
analysis of data produced at SPS and RHIC\cite{Lisa:2005dd}
The second term of \eq{pwa} is written to show that the 
emission function is determined by  the product  of a plane wave factor
and the current that emits the pions. This is the standard result of
scattering theory, if final state interactions are ignored.
Here and throughout the paper we use natural units in which 
 $\hbar$ and $c$ are unity. 
Our use of the subscript 0 denotes that
the effects of  final state interactions are ignored.

Next we discuss the relationship between the function $S_0(x,p)$ and the
single-pion production cross section. If the source current couples linearly 
with the pion field (with a chosen  interaction Lagrangian, $L_I$), and no final state interactions  occur,
the matrix element of the interaction $-L_I$   between the 
vacuum and single-pion final state is  
\bea
&&\widetilde {J}(p)\equiv \int d^4x e^{ip\cdot x}J(x)\;=\;\int d^4x \phi_\bfp(x)J(x),\label{pwft}\eea
where $\phi_\bfp=e^{ip\cdot x}$ is the solution of 
\bea
\left({\partial^2 \over \partial t^2}-\nabla^2+m_\pi^2\right)\phi_\bfp= 0.\label{freeq}\eea
In this plane wave (pw) approximation, the covariant   single-particle
 emission function  can be expressed as
\bea
&&{\cal P}_1^{\rm pw}(p)= \langle  \big\vert\widetilde{J}(p)\big\vert^2
\rangle=\int d^4xS_0(x,p), \eea 

The literature \cite{Wiedemann:1999qn}
presents the two-particle emission function as 
\bea &&
S_0(x,K)=\int 
{d^4y\over2(2\pi)^3}\exp{(-i K\cdot y)}\langle {J}^*(x+y/2) {J}(x-y/2)\rangle.\label{s000}\eea
The quantity that enters in the HBT correlation is $S_0(x,K,q)$ with 
\bea S_0(x,K,q)\equiv S_0(x,K)e^{-iq\cdot x}
=\int 
{d^4y\over2(2\pi)^3}\langle e^{-ip_1\cdot(x+y/2)} {J}^*(x+y/2) 
 e^{ip_2\cdot(x-y/2)}{J}(x-y/2)\rangle\equiv S_0(x,p_1,p_2),
 \label{pwa2}
\eea
where $p_1,p_2$ are   the four-momenta of   the two   on-shell emitted pions,
and
\bea q=p_1-p_2,\; K={1\over 2}(p_1+p_2).\eea
 The brackets again indicate that one takes an  
 ensemble average over chaotic sources. The second form, \eq{pwa2} shows again that
products of  plane-wave functions and currents determine the observables.
The notation  $S_0(x,p_1,p_2)$ is introduced to emphasize the two-particle nature of the
emission function.

In plane wave approximation the correlation function is given by the expression
\bea &&{\cal P}_1^{\rm pw}(p_1){\cal P}_1^{\rm pw}(p_2)C_0(p_1,p_2)=\langle\left\vert  \int {d^4x_1\over 2 (2\pi)^3}
\int {d^4x_2\over 2 (2\pi)^3}{1\over \sqrt{2}}
\left(e^{ip_1\cdot x_1}e^{ip_2\cdot x_2}+e^{ip_2\cdot x_1}e^{ip_1\cdot x_2}\right)J(x_1)J(x_2) \right\vert^2\rangle,
\label{term1}
\eea
in which the effects of the Bose statistics is explicit. One proceeds by taking the
absolute square to find that the expression contains
four terms. The  direct terms involve 
$ \left(e^{ip_1\cdot x_1}e^{-ip_1\cdot x_1'}\langle J(x_1)J^*(x_1')\rangle 
 e^{ip_2\cdot x_2}e^{-ip_2\cdot x_2'}\langle J(x_2)J^*(x_2)\rangle 
\;{\rm plus}\; 1\leftrightarrow2 \right)$ and the exchange terms involve
$ \left(e^{ip_1\cdot x_1}e^{-ip_2\cdot x_1'}\langle J(x_1)J^*(x_1') )\rangle
 e^{ip_2\cdot x_2}e^{-ip_1\cdot x_2'}\langle J(x_2)J^*(x_2) )\rangle
\;{\rm plus}\; 1\leftrightarrow2 \right).$ The placement of the brackets
arises from the use of chaotic sources: the effects of pions produced by two different chaotic sources average
 to 0\cite{ed73,GKW79}.
  Using the 
absolute square and the previous definitions in \eq{term1} 
yields
the result
\bea
&&C_0(p_1,p_2)=1+\frac{\left\vert\int d^4x S_0(x,p_2,p_1)\right\vert^2}
{ \langle  \big\vert\widetilde{J}(p_1)\big\vert^2\rangle \langle  \big\vert\widetilde{J}(p_2)\big\vert^2\rangle}.
\label{c0pw}
\eea

\section{Introducing Final state interactions}
\label{sec:fsi}
Our emission function, $S_0$, is not meant to be the same as that obtained
in conventional hydrodynamic calculations of hadronic spectral functions.
In such calculations\cite{Wiedemann:1999qn} 
the quantity $S_0(x,K)$ is assumed to be  a local equilibrium
Bose-Einstein distribution localized on a 3-dimensional freeze-out hypersurface that
separates the thermalized interior of the hot dense medium  from the free-streaming
particles on its exterior\cite{Cooper:1974mv}. Here we assume that the emitted pions
(and other hadrons) undergo significant
interactions  while escaping the hot dense medium. The emitted
pions interact both with the  hot dense matter and other hadrons as they escape the
system. This is emission from a coexistence  phase.   
The purpose of the present section 
is to provide a formalism that allows such effects to take place, while also
including the usual effects of emission from a freeze out surface.

\subsection{Distorted Waves}
We wish to include the effect that 
an escaping pion has to "fight" its way through the medium. Our formalism relies heavily
on earlier derivations in Ref.~\cite{GKW79}.
The effects of the fighting distort the wave away from its plane wave form dictated by \eq{freeq}.
These interactions of the pions with the medium,
 represented by the optical potential $U$,
lead to a modified equation (approximated as   a one-body equation)
 \bea
\left({\partial^2 \over \partial t^2}-\nabla^2+U+ m_\pi^2\right)\psi= 0.\label{h2}\eea
This equation provides an 
approximate treatment of the complicated final state interactions.
In principle, the optical potential should be related to the underlying pion source currents.
We adopt a phenomenological approach. 
The resulting S-matrix depends on the exact solution,  $\psi_\bfp^{(-)}(x)$,
to this equation with the out boundary condition
of approaching at $t\to\infty$, the free wave $\phi_\bfp(x)$\cite{BJD}. 

Ref.~\cite{GKW79}
uses standard S-matrix reduction techniques to show that matrix element of the pion field 
 between the vacuum and a one-pion final state $p$ is simply the complex conjugate of  
$\psi_\bfp^{(-)}(x)$, if 
 the optical potential
is introduced as in \eq{h2}. This means
 that in calculating the matrix elements of $-L_I$ between
the vacuum and multi-pion final states 
one simply replaces the plane wave solutions of 
 \eq{freeq} by the out-going wave solutions of 
\eq{h2}. 
Thus the simple
replacement:
\bea
&&  e^{ip\cdot x}\to  \psi_\bfp^{(-)*}(x)
\label{rg}\eea
is sufficient to include  final state interaction effects within the optical model approximation.
Therefore the single-particle emission function
is given by 
\bea 
S(x,p_1)=\int 
{d^4y\over2(2\pi)^3}
\langle {J}^*(x+y/2) {J}(x-y/2)\rangle
\psi_{\bfp_1}^{(-)}(x+y/2) 
\psi_{\bfp_1}^{(-)*}(x-y/2).
\eea 
This expression reduces to the plane wave form $S_0(x,p_1)$ \eq{s000} if the distorted wave $\psi_{\bfp_1}$ are
replaced  by plane waves. 
The two-particle 
emission function that includes the effects of final state interactions is obtained by
replacing \eq{s000}  by the result  
\bea  S(x,p_2,p_1)=
S(x,K,q)=\int 
{d^4y\over2(2\pi)^3}
\langle {J}^*(x+y/2) {J}(x-y/2)\rangle
\psi_{\bfp_1}^{(-)}(x+y/2) 
\psi_{\bfp_2}^{(-)*}(x-y/2), \label{dwa}
\eea
which applies for calculating the two-particle emission function. In the plane wave
limit, $S(x,K,q)\to S_0(x,K)e^{-iq\cdot x}.$

Physical observables for the emission of two pions of momenta $p_1,p_2$
are determined by the correlation function
$C(q,K)\quad (K={1\over2}(p_1+p_2),\;q=p_1-p_2)$, which is given by
\bea C(q,K)=1+\frac{\vert\int d^4x S(x,K,q)
\vert^2}{\int d^4x S(x,p_1)
\int d^4x S(x,p_2)}.
\label{4d}\eea One obtains 
 the usual expression (\ref{c0pw}) if the wave functions
$\psi_\bfp^{(-)}$ are replaced by plane 
wave functions ($\psi_\bfp^{(-)}(x)\to e^{-ip\cdot x}$).

The expression (\ref{dwa}) contains the ensemble average of the currents.
This may be expressed in terms of
$S_0(x,K)$ by taking the Fourier transform of Eq.~(\ref{pwa}).
 We then obtain the convolution formula:
\bea && S(x,K,q)=\int d^4K' S_0(x,K')
\int {d^4y\over(2\pi)^4}\; e^{i K'\cdot y}
\psi_{\bfp_1}^{(-)}(x+y/2) 
\psi_{\bfp_2}^{(-)*}(x-y/2)\label{convo}
,\eea
where the subscripts indicate the momenta $p_1,p_2$ of the detected pions
and $K={1\over2}(p_1+p_2)$.
The quantity $S(x,K,q)$ is used to compute  experimental observables in the same way
that $S_0$ was previously used.
 This is the distorted wave emission function DWEF formalism.
The expression (\ref{convo}) is the coordinate space version of Eq.~(5.25)
 of Ref.~\cite{GKW79}.
 
We emphasize that  $S_0$ (which reflects the
 true properties of the source) is no longer directly related
to observables--the appearance of distorted waves obscures the relationship between
the data and the true properties of the source.

\subsection{Extracting HBT radii}
\label{extract}
The correlation function of \eq{4d}  are related to HBT radii    
in two different ways. 
In the first method, one treats the momentum differences $q_{O,S,L}$
 as  small quantities and then expands keeping
terms to second order so that: \bea
C(q,K)-1\approx 1-q_O^2R_O^2-q_S^2R_S^2-q_L^2 R_L^2,\label{cqk}\eea
where $q_O$ is the transverse component that is parallel to the direction of $\bfK$, 
$q_S$  is the transverse component that is perpendicular to the direction of $\bfK$,
and $q_L$ is the longitudinal component.  Here and below, because we are focusing on central collisions we ignore the $q_O q_L R_{OL}^2$ cross term.  Another parameterization is 
 \bea
C(q,K)-1\approx \exp (-q_O^2R_O^2-q_S^2R_S^2-q_L^2 R_L^2),\label{ckqgauss}\eea

In practice, describing data and extracting
radii require using: \bea
&&C(q,K)-1\approx \lambda\exp (-q_O^2R_O^2-q_S^2R_S^2-q_L^2 R_L^2),\label{ckqgauss1}\\
&&C(q,K)-1\approx \lambda(1-q_O^2R_O^2-q_S^2R_S^2-q_L^2 R_L^2).\label{ckqquad}\eea
Here the reduction factor $\lambda$ (typically about 1/2) is the fraction of pairs that originate in the space-time region relevant for correlations,  see the review
 \cite{Lisa:2005dd}. Current understanding \cite{Lisa:2005dd}
is that the HBT data are consistent with incoherent
emission, and  accounting for the many  pions that  are
produced by the decays of resonances far outside the collision region
can reproduce the factor $\lambda.$  These ``halo'' pions do not have
a BE-enhanced correlation in the $q$ region measured
with the pions emitted from the core of hot dense matter, but they cannot be experimentally separated from the latter.

Our calculations employ the core-halo model \cite{core-halo} in which
pions are assumed to arise either from the hot, dense 
core or from the halo.   This model is
a simplified version of more detailed treatments of resonance decays that are
discussed in the review \cite{Wiedemann:1999qn}.


The influence of resonances affects the extraction of radii and the measurements of
the pion spectrum in different ways. 
We account for this in our phenomenological analysis, see the erratum of 
Ref.~\cite{Cramer:2004ih} and Sect. IX-A.  Here we note only that
effects of  pions produced by short-lived resonances (such as $\Delta, \rho,\cdots$)
 that are not explicitly
included in the pionic wave equation are included in  the function 
$S_0$. The pions resulting from 
long lived resonances are not included in $S_0$. Pions 
resulting from the decay  of slowly moving $\Omega$ mesons that occur inside
the dense matter system are included in $S_0$, but the pions from rapidly
moving $\omega$ mesons that decay outside of the system are not
included in $S_0$.

The approximate forms (\ref{ckqgauss1},\ref{ckqquad}) provide two
 ways to extract radii. The former suggests that 
\bea R_i^2={1\over q_i^2}\ln{\lambda\over C(q_i,K)-1},\quad i=O,S,L\label{logform}
\eea while the latter implies
\bea R_i^2={1\over q_i^2}{C(q_i,K)- 1\over\lambda},\quad i=O,S,L\label{quadform}
\eea
Eq.~(\ref{logform}) can be used for values of $q_I$ such 
that  $(C-1)/\lambda$ can be approximated by 
a Gaussian function. The use of Eq.~(\ref{ckqquad}) 
 requires that $q_iR_i\ll 1$. We show below in Sect.~\ref{sec:corr} that,  while
the correlation functions are not  Gaussians (so that the squared radii are not moments of the
correlation function), the Gaussian parameterization
is quite accurate at the modest (but not very small) values of $q_i\sim 30$ MeV/c
that dominate the experimental extraction of radii.

\section{Symmetries of $S_0(x,K)$ and the form of the pion distorted waves}
\label{sec:sss}

 We use the hydrodynamic parameterization of the source
of Ref.~\cite{CL96a,H96,Tomasik:1997eq}. Generally based on the
Bjorken tube model, it is  given by 
\begin{eqnarray}
   S_0(x,K) \, d^4x & = & {M_\perp \cosh(\eta-Y) \over
            (2\pi)^3 }
        \; {1\over\exp \left[ {(K \cdot u(x)-\mu_\pi) \over T(x)}\right]-1}
        \; \rho(b)\exp \left[ 
            - {{(\eta- \eta_0)}^2 \over 2 (\Delta \eta)^2}
           \right]         \nonumber \\ 
&  & \times 
            {\tau \, d\tau}\left[\frac{1}{\sqrt{2\pi(\Delta \tau)^2}}  \;
            \exp \left(- {(\tau-\tau_0)^2 \over 2(\Delta \tau)^2}
           \right )\right] \, d\eta \, b\, db\, d\phi,
\label{3.1}\\
d^4x&=&\tau d\tau\;d\eta\;b\;db\;d\phi.
\end{eqnarray}
Here, $\mu_\pi$ is the pion chemical potential, 
 the variables $(b,\phi)=\bfb$ are equivalent to
$\bfx_\perp$,$\;\eta={1\over 2}\ln{t+z\over t-z},\;\tau=\sqrt{t^2-z^2}, M_T=\sqrt{K_\perp^2+m_\pi^2},
Y={1\over2}\ln {E_K+K_z\over E_K-K_z}.$ 
The factor in the brackets involving $\tau$ has the same normalization as
the delta function: $\delta(\tau-\tau_0)$. We use a 
 Bose-Einstein distribution instead of
the  Boltzmann distribution of Ref.~\cite{Tomasik:1997eq} and
 also allow the transverse density $\rho(b) $ to have  a general
form instead of a Gaussian. 
Note that the model contains the parameters: $R,\eta_0,\Delta \eta,\Delta\tau,
\tau_0,\eta_f$. As stated originally,    temperature 
gradients were included, but we treat the temperature
 as a constant for our numerical calculations. 
We shall see below that our formalism can easily be generalized to allow 
 the temperature to
vary as a  function of $b$.  

 The  longitudinal $\eta,\tau$
 and transverse
$(\bfb, \bfK)$ variables can be separated in the following form:
\bea
S_0(x,K)&=&{\cal S}_0(\eta,\tau,Y)\;B_\eta(\bfb, \bfK)\label{fact}\\
B_\eta(\bfb, \bfK)&\equiv& {M_\perp\over \exp [(K\cdot u-\mu_\pi)/T]-1}
\rho(b)\label{bigb}\\
{\cal S}_0(\eta,\tau,Y)&\equiv &
{\cosh(\eta-Y) \over
            (2\pi)^3 }
                                \; \exp \left[ 
            - {{\eta}^2 \over 2 (\Delta \eta)^2}
           \right]\frac{1}{\sqrt{2\pi(\Delta \tau)^2}}  \;
            \exp \left[- {(\tau-\tau_0)^2 \over 2(\Delta \tau)^2}
           \right ].     
\label{emit}\eea 
Here $\rho(b) $ is a  function, normalized as $\rho(0)=1$ 
that represents the transverse density. To be specific we use 
\bea\rho(b)= [1/(\exp((b-R_{WS})/a_{WS})+1)]^2.\label{rhodef}\eea 
This distribution has a correct exponential fall-off at large values of $b$, and
different choices of the  parameters $R_{WS},\;a_{WS}$ allow a variety of shapes
to be assumed.

We concentrate on the kinematics of the STAR experiment which detects pions 
within  half a unit of rapidity of moving perpendicular to the beam, so
we take $Y=0$. This means that the average momentum is transverse: $\bfK=\bfK_\perp=\bfK_T$.
 With colliding beams of equal mass and energy
there is a fore-aft symmetry along the longitudinal axis, so
that we use 
       $\eta_0=0.$
Therefore it is
useful to define
\bea
{\cal S}_0(\eta,\tau)\equiv {\cal S}_0(\eta,\tau,Y=0).\eea

The velocity field $u(x)$ describing  the dynamics of the expanding source 
is  parameterized  by \cite{WSH96}
\begin{equation}
 \label{3.3}
   u^{\mu}(x) = \left( \cosh \eta \cosh \eta_t(b), \,
                     \cos \phi \, \sinh \eta_t(b),  \,
                     \sin \phi \, \sinh \eta_t(b),  \,
                     \sinh \eta \cosh \eta_t(b) \right) ,
 \label{umu}\end{equation}
with $\phi$ the angle between $\bfK_\perp$ and $\bfu$ (or when appearing in 
 the single-particle emission function, it is the angle between $\bfp_i$ and $\bfu$). Eq.~(\ref{umu})
implements a boost-invariant longitudinal flow profile
$v_L = z/t$, with a linear radial profile of strength $\eta_f$
for the transverse flow rapidity:
 \begin{equation}
 \label{3.4}
  \eta_t(b) = \eta_f {b \over R_{WS}} \, .
 \end{equation}
The exponent that enters in the Bose-Einstein distribution 
 is given by the four-vector dot product:
  \begin{equation}
    \label{3.5}
    K\cdot u(x) = M_\perp \cosh(\eta - Y) \cosh \eta_t(b) 
                - K_\perp \, \sinh \eta_t(b) \,\cos \phi \,.
  \end{equation}
The presence of a non-vanishing value of  $\eta_f $ 
causes the emission function to
depend on both $K$ and 
$M_\perp$.
The Bose-Einstein distribution of Eq.~(\ref{bigb})
is evaluated  as a sum of Boltzmann distributions:
\bea &&{1\over \exp \left[ {K \cdot u(x)- \mu_\pi\over T}\right]-1}=\sum_{n=1}^\infty
\exp({-K\cdot u +\mu_\pi\over T_n}),\;T_n\equiv {T\over n}.\label{sumbe}\eea

The parameterization (\ref{3.1}) is motivated 
by hydrodynamical models with approximately boost-invariant longitudinal 
dynamics. It uses thermodynamic and hydrodynamic parameters and 
appropriate coordinates. 
The ``emission function'' given above was originally 
intended to parameterize the distribution of points of last 
interaction in the source.  In conventional treatments, the Cooper-Frye\cite{Cooper:1974mv}
matching
procedure is used to obtain distributions of detected particles. Our approach is more general. 
We assume that 
pions can be formed at any space-time point (including but not limited to the freeze-out surface)
during the collision, and propagate through   the dense medium while interacting 
 before being detected.
 Thus we take $S_0(x,K) $ to be the emission function in the absence of final
state interactions.

We note that the 
 emission function of Eq.~(\ref{3.1}) has been previously used in the ``blast-wave model''
 \cite{Retiere:2003kf}, and we would like to comment on the differences between our formalism and that
model. These include:
\begin{itemize}
\item The blast-wave model does not attempt to reproduce the normalization of the spectrum, and sets 
the chemical potential $\mu_\pi$ to 0. We find that taking $\mu_\pi$ to be the pion mass allows us to reproduce 
 both the normalization and the {\it shape} of the pion momentum spectrum.

\item The blast-wave model uses the smoothness approximation   and computes
HBT radii as moments of the emission function.  

\item The blast-wave model uses plane waves and therefore omits the effects of the optical potential.

\item The blast-wave model uses a parameterized emission function to describe the six-dimensional distribution of pions on some freeze-out hypersurface, while the emission function of the DWEF model describes the initial emission of pions within the hyper-volume of the hot, dense medium produced by the collision.  Therefore, it is not appropriate to compare emission-function parameters like $T$ and $R$ that are derived from blast-wave fits with those of the present model.

\end{itemize}

\section{Final state interactions and the optical potential}
\label{sec:fsiop}

The salient feature 
of  the 200  GeV data is the high density of the produced matter, so we
treat  the effects of pion final state 
interactions with a dense medium.
We adopt  a single-channel  approach that 
uses 
the  
interaction--distorted incoming wave
$\Psi_{\bfp_1}^{(-)*}(x_1)$. 

 We assume that the 
matter produced in the central region of the collision 
is cylindrically symmetric with a very long axis, so that an expression of the form
(\ref{fact}) is valid. In that case, the optical potential $U$ representing
the  interaction between a pion and the medium is a 
complex, azimuthally-symmetric function 
depending on pion momentum and local density.  Within our
formalism the influence of some time-dependent effects in $U$ introduced by
the time-dependent source $S_0$ is incorporated in the energy dependence of
the optical potential, and  the pion-medium interaction
time is %
restricted by $S_0$.

The optical potential accounts for the interaction between each pion and the surrounding medium,
but does not include the interaction between the two pions. In particular, 
the Coulomb interaction is known to be important. The experimental analysis removes this effect
before extracting radii from the data. 
The Coulomb interaction between pions is of long range and 
its important effects occur when 
the  pions are  outside the medium. Indeed, specific analyzes show
that Coulomb effect occur only at very low relative momenta \cite{STARHBT}.
In contrast,
  the optical potential acts only when pions  are inside.
We therefore expect that  presence 
of the optical potential would not influence the removal of
the Coulomb interaction. A quantitative treatment of the effect of the optical potential
on the removal of Coulomb effects 
would involve solving a three-body problem. This is not warranted at
the present stage of development, but might eventually become worthwhile.

The optical potential accounts for 
situations in which the pion changes energy or disappears entirely due to its
interactions with the dense medium. We do not assume to  know the content
of the dense medium, and therefore  will use a phenomenological optical potential.
 But it is worthwhile to consider a simple example to get an idea
about how large the optical potential can be.
Suppose, {\it e.g.}, that 
the medium is a gas of pions. Then 
$\pi\pi$ scattering would be  the origin of $U$.
In the impulse approximation, the central optical potential would be
$U_0=-4\pi f\rho_0$, where $f$ is the complex 
forward scattering amplitude and $\rho_0$ the central density. 
For low energy pion-pion interactions, $4\pi$Im$[f(p)]=p\sigma,$
with $\sigma\approx 1\; $mb. At a 
momentum $p=1~\rm{fm}^{-1}=197.3 $ MeV/c, 
using a pion density about
ten times the baryon density of ordinary nuclear matter, 
Im$[U(0)]\approx -0.15~\rm{fm}^{-2}$, representing significant opacity.

The optical potential must be an analytic function of energy, and therefore the
existence of an imaginary part mandates the existence of a real part. Thus, any
analysis needs to treat $U$ as a complex function. Under certain circumstances
the real part can be very large. For example,
 if two interacting 
pions each have less energy than half  of the rho meson mass, the 
final state interactions caused by virtual transitions to a rho meson
would be strongly attractive. Additionally, the influence of chiral symmetry restoration
can lead to a strong real part. This is discussed next.

\subsection{Chiral Symmetry Restoration}
\label{sec:csr}

Suppose the dense medium is one in which chiral symmetry is restored. 
This means that the value of the quark condensate vanishes, an effect that
could be caused by an 
 increase in temperature or density. The pion mass 
is proportional to the quark condensate via the GMOR relation \cite{Gell-Mann:1968rz}
\bea 
m_\pi^2f_\pi^2=-{m_u+m_d\over2}\langle0\vert\bar{u}u +\bar{d}d\vert 0\rangle
,\eea
where $f_\pi$ is the weak pion decay constant $\approx 93 $ MeV.
It is believed that $m_\pi,f_\pi$ and the  condensate 
$\langle0\vert\bar{u}u +\bar{d}d\vert 0\rangle$ all depend on temperature
and density. If one takes the Brown-Rho\cite{Brown:1991kk} 
scaling relation for $f_\pi$ and
the perturbatively calculated temperature dependence of the condensate, the
pion mass is proportional to the cube root of the condensate, and
 therefore vanishes for sufficiently large temperatures. 
See the reviews \cite{Koch:1996vt}.
Suppose the 
optical potential arises {\it only} from the temperature dependence of the pion mass.
Then the  Klein-Gordon equation would take the form:
\bea (-\nabla^2 + m_\pi^2(T))\psi = (p^2+m_\pi^2)\psi,\label{vk}
\eea
 for regions inside the medium. The effects of the medium are incorporated through
the difference between $m-\pi^2(T)$ and  $m_\pi^2$. If one re-writes \eq{vk} as a Klein-Gordon
equation $ (-\nabla^2 + U)\psi = p^2\psi,$ then the optical potential takes the form:
\bea
U(b)=(m_\pi^2(T)-m_\pi^2)\rho(b),\label{opt2}\eea
in which the finite extent of the medium is accounted for by the factor $\rho(b)$.
If $m_\pi(T)$ approaches zero, the optical potential is attractive with magnitude $m_\pi^2.$

A more recent study by Son \& Stephenov\cite{Son:2002ci} provides a more detailed  
treatment of the effects of chiral restoration in which 
 the general $p$-wave nature of the low-energy interaction between
pions and any target is included. We are guided by this work.
 Son \& Stephenov use the dispersion relation for low
momentum pions in infinite nuclear matter\cite{boy,Son:2002ci}:
\bea
\omega^2=u_\pi^2(\hat{p}^2 +m_\pi(T)^2),\label{mode}\eea
where $\hat{p}^2$ is the infinite-sized matter version of $-\nabla^2$.
The quantity $u_\pi$ is termed the pion velocity, even though it is only that
when $m_\pi(T)$ vanishes. The term  $m_\pi(T)$ 
is denoted the pion screening mass. This quantity appears in the expression for
the static Euclidean pion correlator\cite{Son:2002ci}.
The energy of a pion at $\bfp=0$ is termed the pion pole mass.
  The free pion mass is $m_\pi$. In Ref.~\cite{Son:2002ci} \eq{mode} applies only
for $T<T_c$. For larger temperatures, 
chiral symmetry is restored and pions are massless.

Defining $t\equiv (T_c-T)/T_c$, 
SS find $m_\pi^2(T)\sim t^{\beta-\nu},\;u^2\sim t^\beta$, with $\beta<\nu$, e.g
$\nu=.73,\beta=.38$\cite{Baker:1977hp}. These equations 
are valid for temperature close to (but not too close to) the critical point.  
Another view about the dispersion relation can be found in  Ref.~\cite{sas}.
This general discussion about the influence of chiral restoration
provides some guidance, 
but does not tell us exactly to use.

We wish to obtain
an equivalent optical potential and see if it is attractive or repulsive.
Use  the Klein Gordon equation in the form
\bea
\hat{p}^2 +U+m_\pi^2=p^2+m_\pi^2 .\label{opt}\eea
In regions
outside the dense medium  where $U=0$, the operator $\hat{p}^2$ is simply the square of the
momentum, $p^2$. 
Subtract (\ref{opt}) from (\ref{mode}) to obtain
\bea U=u_\pi^2m_\pi^2(T)-m_\pi^2 +(u_\pi^2-1)\hat{p}^2,\label{uopt3}\eea
an expression that is the sum of two negative definite terms.
This form can be simplified by using the wave equation (\ref{opt}) to
remove term $\hat{p}^2$. Then one finds a momentum-dependent optical potential:
\bea
 U={u_\pi^2m_\pi^2(T)-m_\pi^2 +(u_\pi^2-1)p^2\over u_\pi^2}.\label{uss}\eea
Note that if $u_\pi$ becomes really small the optical potential becomes 
very strongly attractive.

For matter of finite size, the term
 $\hat{p}^2$ can also be  
interpreted  as 
 $-\nabla\cdot c\nabla$ which is the Kisslinger term \cite{lsk}. For infinite 
nuclear matter  only forward 
scattering occurs and the two terms are identical, but 
 differences may 
arise for scattering from media of finite size. 
One can not tell the difference between the  
two terms at the  start, so that  we find
a general form
\bea
U(b)=- (w_0 + w_2 (1-\epsilon)    p^2)\rho(b)-\epsilon w_2\nabla \rho(b)\cdot\nabla,\label{u526}\eea
with  both the real and imaginary parts of $w_0,w_2,\epsilon$ 
positive  for attractive interactions, and $\rho(b)$ taken from \eq{rhodef}. 
This simple form is strictly valid only for  low-energy pions.
The limit of an infinite tube is used to write $U$ as independent of $z$.
We find for the 200 GeV data, 
that for temperatures below our assumed value of 
193 MeV the fitting prefers a very small value of the Kisslinger term, so we
simply set $\epsilon = 0$. 
We also take 
$w_0$ real (so that there is no opacity at p=0).  
We shall see  below, that if the temperature is 
set to a value greater than 193 MeV, the fit is improved 
by including the gradient terms at about the 20\% level
 ($\epsilon \approx 0.2$). The simple form \eq{u526} is sufficient to account
for the data we study, $K_T\leq 600$ MeV/c, but we remind the reader that
the interaction strength 
does not grow as 
$p^2$ for $p$ much greater than about 400 MeV/c.

We shall discuss    the precise parameters 
of the optical  potential in
subsequent sections. For now we may simply proceed using the assumption that 
$U$ is a complex, azimuthally-symmetric function 
depending on pion momentum and local density of the form of \eq{u526}.

\section {Finding $\psi^{(\pm)}({\bf x})$}
\label{sec:psi}

The evaluation of the emission    function (\ref{convo}) requires performing an eight dimensional integral using
a distorted wave. 
We shall use symmetries to reduce the number of numerical evaluations. Doing this depends on obtaining a compact
expression for the distorted wave.
 In the present section, we show how the distorted waves are evaluated.

The first step is to realize that the function $\psi_\bfp^{(-)}(x)$  
represents an energy-eigenfunction\cite{GKW79} provided the optical potential does not
change with time. We shall show below that the value of $\delta \tau$ is not large. Thus the
the time-independent optical potential that we use can be thought of as
a time-averaged optical potential.
 So we have
\bea \psi_\bfp^{(-)}(x)=e^{-i \omega_p\;x^0}\Psi_\bfp^{(-)}({\bf x}).\label{eig}\eea

To proceed we need to examine the properties of the wave function
$\psi_\bfp^{(-)}({\bf x})$.
It is conventional to compute $\psi_\bfp^{(+)}({\bf x})$, and we will follow
this convention. Then we  use time reversal invariance in the form
\bea
\Psi_\bfp^{(-)}({\bf x})={\Psi_{-\bfp}^{(+)}}^*({\bf x}) 
\label{pm}\eea
to obtain the desired wave function.

The next step is to realize that for central collisions,
$Y=0$,  the emission function (\ref{emit}) has a cylindrical symmetry. This means that the  expected optical
potential is azimuthally symmetric. If we take the matter to have the form of a very long tube, the optical
potential will be independent of $z$. Then one obtains 
a solution that takes a product form
\bea
\Psi_{\bfp_{1,2}}^{(-)}({\bf x})& = &
e^{\mp iq_L z/2}\psi_{\bfp_{1,2}}^{(-)}({{\bf x_\perp}=\bfb}),\label{psidef}\\
\quad \bfp_{1,2} & = & \bfK\pm \bfq/2\pm {\bf\widehat{z}}\;q_L/2, \label{waver}
\eea 
where the vector  $\bfq$ is defined as a  transverse vector $\bfq\cdot {\bf\widehat{z}}=0$.

We may obtain the wave function $\psi_\bfp^{(+)}({\bfb})$ by solving
the wave equation
\bea \left(-\nabla_\perp^2 +U(b)\right)\psi_\bfp^{(-)*}({\bfb})=p^2
\psi_\bfp^{(-)*}({\bfb}).
\label{wave}\eea
If $U=0, \;\psi_\bfp^{(+)}({\bfb})=e^{i\bfp\cdot\bfb}$.
Many previous treatments of opacity can be  understood as using 
the eikonal approximation to obtain solutions to  Eq.~(\ref{wave}). 

We take the optical potential to have the azimuthally-symmetric form of \eq{u526} so that
the solution for $\psi_\bfp^{(\pm)}(\bfb)$ 
can be expanded in partial wave
form in plane polar coordinates ($b\equiv\sqrt{\bfx_\perp^2},{\phi},\cos\phi\equiv
\hat{\bfp}\cdot \hat{\bfb}$):
\bea
&&\psi_\bfp^{(+)}({\bfb})=\sum_{m=-\infty,\infty}f_m(p,b)i^m\;e^{im\phi},\\
&&\psi_\bfp^{(+)}({\bfb})=f_0(p,b)+2\sum_{m=1,\infty}f_m(p,b)i^m\;\cos{m\phi},
\label{pw}\eea
with (\ref{pw}) taking into account the invariance of the differential equation
for $f_m$ under the interchange $b\to-b$. 
Note that we may  use Eq.~(\ref{pm}) to find
\bea
\psi_\bfp^{(-)*}({\bf x_\perp})=f_0(p,b)+2\sum_{m=1,\infty}f_m(p,b){(-i)}^m\;\cos{m\phi}
.\label{psim}\eea
In practice a finite number of
terms is needed, with $m\le m_{max}\approx 2pR/\hbar$.

  Use Eq.~(\ref{pw}) in (\ref{wave}) to find
\bea\left( {d^2\over db^2} +{1\over b} {d\over db}+(p^2- {m^2\over b^2})\right)
f_m(p,b)
-U({b^2\over R^2})f_m(p,b)=0.\label{diff}\eea
Note that for large enough $b$, $U$ vanishes and the $f_m$ are linear
combinations of Bessel $J_m$ and Neumann $N_m$ functions or Hankel functions $H_m^{(1,2)}$. 

This differential equation can be solved numerically using the Runge-Kutta
technique. One determines the wave function by matching the numerical
solutions to the analytic solution
\bea
f_m(p,b) =A_m\left( J_m(pb)+T_m H_m^{(1)}(pb)\right).
\eea

One matches the numerical function and its derivative
(at large enough $b$ so that $U=0$) so as to determine the constants $A_m,T_m$.
The normalization of $\psi_\bfp^{(\pm)}$
 is such that $A_m=1$--asymptotically the wave is a sum of an ordinary plane wave and
an outgoing wave, and the functions $f_m(p,b)$ can be regarded as phase shifted 
Bessel functions.  Using
the  partial-wave form of two-dimensional wave function (\ref{pw}) will simplify the 
evaluation of $S(x,K)$ of Eq.~(\ref{convo}).

\section {The Distorted 
Wave Emission Function (DWEF) and the Large Source Approximation (LSA) }\label{sec:dwef}

 Using (\ref{eig})  in
Eq.(\ref{convo}) allows the integrals over $y^0$ and $K^{0'}$ to be
evaluated  as:
\bea && S(x,K,q)=\int {d^3K'\over(2\pi)^3} S_0(x;K^0,\bfK')
e^{i(\omega_2-\omega_1)x^0}
\int d^3y\; e^{-i\bfK'\cdot \bfy}
\Psi_{\bfp_1}^{(-)}(\bfx+\bfy/2) 
\Psi_{\bfp_2}^{(-)*}(\bfx-\bfy/2),\label{dwa0}\eea
and
using (\ref{waver}) allows the integrals over $y^3$ and ${K^3}'$ to be  evaluated
yielding a four-dimensional integral:
\bea
S(x,K,q) & = & {1\over (2\pi)^2}{\cal S}_0(\tau,\eta) e^{iq^0t-iq_l z}
\int d^2b' \widetilde{B}_\eta(\bfb,\bfb')
 \psi_{\bfp_1}^{(-)}(\bfb      +\bfb'     /2) 
\psi_{\bfp_2}^{(-)*}(\bfb      -\bfb'     /2),\label{dwa1}\eea
$$
\widetilde{B}_\eta(\bfb,\bfb')\equiv \int d^2K'_{T}\;B_\eta(\bfb      ,\bfK'_{T}     )
\exp\left[-i\bfK_T'\cdot\bfb'     \right]. $$

The result (\ref{dwa1}) still requires the  evaluation of  an 6-dimensional 
integral (over $\tau,\eta,\bfb',\bfK'_T$) to obtain the correlation function. We search for 
 simplifications.
The integral (\ref{dwa1}) simplifies if we ignore the effects of transverse flow
rapidity. So to gain insight,  let's set 
\bea \eta_f=0,\eea   consider a fixed value of $\eta$, and take one of the terms in 
the series (\ref{sumbe}). Then 
\bea
&& \widetilde{B}_\eta(\bfb,\bfb')=\rho(b)
\;g(\bfb'^2)\\
&&g(\bfb'^2)=2 \int d^2K_\perp\;M_\perp\exp\left[{- M_\perp \cosh\eta 
               \over T}\right]\;\exp\left[-i\bfK_\perp\cdot\bfb'\right] 
\label{gfun}\eea
Fig.~\ref{figg}  shows $g(b)$ 
for the case $T/\cosh\eta=m_\pi$. It is clear from (\ref{gfun}) that the quantity $T$ controls the 
range of allowed values of $K_\perp$, so that the extent 
of $b'$ is of order $1/T\approx 1 \;{\rm fm}$ for $T$ = 200 MeV. This is much smaller
than the size of the presumed fireball that controls the  extent of $b$ in (\ref{dwa1}). Thus it is
natural to think of neglecting the terms involving $\pm\bfb'/2$ of (\ref{dwa1}).
This however,
is too extreme an 
approximation because it would not lead to the formalism of Sect.~1 in the plane wave limit.
 Instead, we use the approximation 
\bea \psi^{(-)}(\bfb\pm\bfb'/2)\approx e^{\mp i\bfp_\perp\cdot\bfb'/2}\psi^{(-)}(\bfb)
\label{LSA1}\eea
This is exact in the plane wave limit, but its wider validity relies on the replacement
\bea 
\psi_{\bfp_i}^{(-)}(\bfb+\bfb'/2) 
\psi_{\bfp_j}^{(-)*}(\bfb-\bfb'/2)\;g(\bfb'^2)\approx
\psi_{\bfp_i}^{(-)}(\bfb)
\psi_{\bfp_j}^{(-)*}(\bfb)\;g(\bfb'^2)\exp{(i\bfK_\perp\cdot\bfb')},\label{LSA}
\eea 
that requires that the size of the source be much larger than  ${n/T}$, where $n$ is the expansion order in the Bose-Einstein distribution (\ref{sumbe}).  
Our sources typically have a diameter of about 25 fm, and $T\sim 1 \;{\rm fm}^{-1}$, so that 
$n$ must be no bigger than about 25. We achieve numerical convergence with 
$n<10$, so that the source is truly large enough for our approximation. We denote
\eq{LSA} to be the ``Large
Source Approximation'' (LSA) and 
 use it 
to  immediately integrate over $\bfb'$ and to obtain a simpler version of (\ref{dwa1}):

\begin{figure}
\includegraphics[width=10 cm]{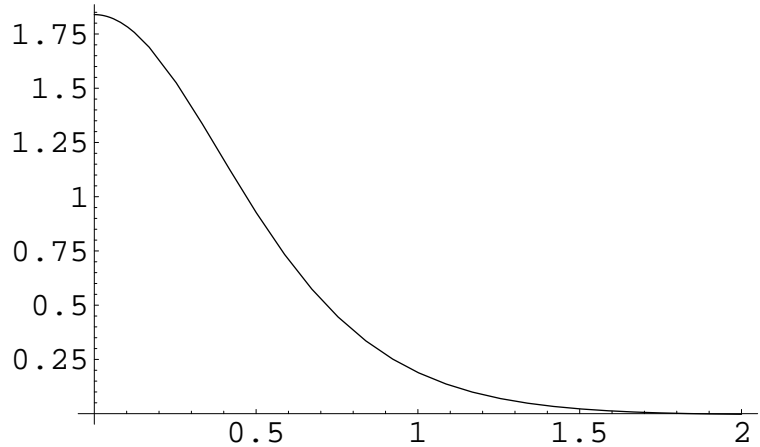}
\caption{\label{figg} Plot of g(b), showing cutoff around $b=1/T$, with $T=1/m_\pi$ and $b$ in fm.} 
\end{figure}

\bea  S(x,K)= {\cal S}_0(\tau,\eta,Y)
e^{i(\omega_2-\omega_1)\tau\cosh \eta}e^{-iq_L\tau\sinh\eta}
 {B}_\eta(\bfb,\bfK)
\psi_{\bfp_1}^{(-)}(\bfb)
\psi_{\bfp_2}^{(-)*}(\bfb), \label{sdefy0}\eea
that is obtained for {\it any} value of
 $\eta_f$. The expansion (\ref{sumbe})
gives
 \bea &&B_\eta(\bfb,\bfK)=\sum_{n=1}^\infty\exp({-K\cdot u +\mu_\pi\over T_n})
M_T\rho(b)\nonumber\\
&&=\sum_{n=1}^\infty \exp({-M_T\cosh\eta\cosh\eta_t(b) +\mu_\pi\over T_n})
\exp({K_T\sinh\eta_t(b)\cos\phi\over T_n})M_T\rho(b),\label{beexp}\eea
 so that Eq.~(\ref{sdefy0}) becomes  
\bea && S(x,K,q)= {\cal S}_0(\tau,\eta,Y)
e^{i(\omega_2-\omega_1)\tau\cosh \eta-iq_L\tau\sinh\eta}
\sum_{n=1}^\infty\exp(-\gamma_n(b)\cosh\eta)B_n(\bfb,\bfK)\psi_{\bfp_1}^{(-)}(\bfb)
\psi_{\bfp_2}^{(-)*}(\bfb)\label{12int}\\&&
B_n(\bfb,\bfK)\equiv\exp(\mu_\pi/ T_n)\exp(K\sinh\eta_t(b)\cos\phi/T_n)\;M_T\rho(b)\label{fulls}\\&&
 S(x,p_i)= {\cal S}_0(\tau,\eta,Y)
\sum_{n=1}^\infty\exp(-\gamma_n(b)\cosh\eta)B_n(\bfb,\bfp_i)\left\vert\psi^{(-)}_{{\bfp}_i}(\bfb)\right\vert^2,\\
&&\gamma_n(b)\equiv {M_T\cosh\eta_t(b)\over T_n}.
\label{sdenom}\eea
 The number of terms in the summation over $n$ 
required to achieve an accurate result depends on the values of $\mu_\pi,\bfK$. 
If desired the quantities  $T,\mu_\pi$ can be treated 
as functions of $b$ \cite{CL96a} in Eq.~(\ref{fulls}), rather than as constants, as we have done here.

\section{Correlation function}
\label{sec:corr}

The specific form of the wave function enables us to express the correlation function as
$C(q_L{\bf\hat{z}}+\bfq,\bfK)$: 
\bea C(q_L{\bf\hat{z}}+\bfq,\bfK)=1+\frac{\vert\int d^4x S(x,K,q)
\vert^2}{\int d^4x S(x,p_1)
\int d^4x S(x,p_2)}\label{cli}\\
\bfp_{1,2}=\bfK\pm q_L/2{\bf\hat{z}\pm \bfq/2}.\label{nicecorr}\eea
The dependence on $q_L$ occurs only in  the numerator.

The remaining task is to perform the integral over $d^4x$. 
We use two separate approaches. The first \cite{Cramer:2004ih} involves
expanding $S(x,K)$, Eq.~(\ref{fulls}), as double power series in $q_L$ and $\tau$, keeping all terms up to second order. 
The second involves exact numerical integration. The two methods yield nearly identical results,
with the second being more accurate and taking only slightly more computer time. We present both
methods here. First, we take $\bfq$ to be small and make expansions Sec.~\ref{subs:expeval}. 
This formalism is applied to obtain numerical results for radii and spectra  in 
Sects.~\ref{subs:num},\ref{subs:auau} and Sect.~\ref{subs:cucu}. 
The formalism to compute the 
correlation functions 
without the use of expansion is contained in  Sec.~\ref{subs:ee}, 
and this formalism is applied to compute correlation functions in Sec.~\ref{subs:cfg}.

\subsection{Evaluation of correlation function by expansion}
\label{subs:expeval}

Now we make the above mentioned expansion keeping terms to order $q_L^2$, (and anticipate
 the integration over $\eta,\tau$).
The term linear
in $q_L$  is an odd function of $\eta$ so that it vanishes when the
 integration over $\eta$ is carried out. Use $\omega_2-\omega_1=-q_o\beta,\beta\equiv K_T/M_T$. 
Thus the expansion of Eq.~(\ref{fulls}) is  
\bea && S(x,K)= 
{\cal S}_0(\tau,\eta,Y)(1-iq_o\beta\tau\cosh\eta-{1\over2}q_o^2\beta^2\tau^2\cosh^2\eta
-{1\over2}q_L^2\tau^2\sinh^2\eta)\non&&\times
\sum_{n=1}^\infty\exp(-\gamma_n(b)\cosh\eta)B_n(\bfb,\bfK_T)
\psi^{(-)}_{{\bfp}_1}(\bfb)\psi^{(-)*}_{{\bfp}_2}(\bfb),\label{2ndo}\eea\\

The effects of the $Y$ dependence of ${\cal S}_0(\tau,\eta,Y)$  that appears for
$q_L\ne0$ is the same for the  numerators and denominators of the correlation functions
and vanish. Hence we do not keep this explicit dependence in the intermediate steps in the
following calculations.

The next step in evaluating  (\ref{cli}) is to  integrate over all 
$\tau, \eta$ using the measure $ d\eta \tau d\tau$.  
The first integral to appear
 arises from the factor of unity appearing inside the parenthesis of (\ref{2ndo}).
It is
\bea
&& I_0\equiv {1\over\sqrt{2\pi}\Delta\tau}\int_{-\infty}^\infty d\eta\cosh\eta\int_{-\infty}^\infty\tau d\tau 
\exp( - {\eta^2 \over 2 (\Delta \eta)^2})
 \exp (- {(\tau-\tau_0)^2 \over 2(\Delta \tau)^2}) \exp(-\gamma_n(b)\cosh\eta)\non
&&\approx2\tau_0\exp({1\over\Delta\eta^2})
K_1(\gamma_n+{1\over \Delta\eta^2})
\equiv \tau_0 f_0(\xi_n)\exp({1\over\Delta\eta^2}) ,\;f_0(\xi_n)=2K_1(\xi_n),
\xi_n\equiv \gamma_n+{1\over \Delta\eta^2}.
\eea
The approximation involves the replacement  \bea\exp( - {\eta^2 \over 2 (\Delta \eta)^2})\to
\exp({1\over\Delta\eta^2})\exp(-{1\over\Delta\eta^2}\cosh\eta).\label{replace}\eea

Similar replacements are made
below. The dominant contributions to the integral  involve small values of $\eta$, so the 
approximation is expected to be very good. The
 appendix shows that the error involved is less
than about 2\%.

We proceed to use the same replacement to evaluate the remaining integrals and find
\bea
&& I_2\equiv{1\over\sqrt{2\pi}\Delta\tau}\int_{-\infty}^\infty \cosh\eta d\eta\int_{-\infty}^\infty\tau d\tau 
\exp( - {\eta^2 \over 2 (\Delta \eta)^2})
 \exp (- {(\tau-\tau_0)^2 \over 2(\Delta \tau)^2})\exp(-\gamma_n(b)\cosh\eta)
(-iq_0\beta)\tau\cosh\eta\non
&& 
\approx (-iq_0\beta)(\tau_0^2+(\Delta\tau)^2)
\exp({1\over\Delta\eta^2})f_2(\xi_n), 
f_2(\xi_n)=K_0(\xi_n)+K_2(\xi_n)=2(K_0(\xi_n)+{K_1(\xi_n)\over\xi_n}),\\
&& I_3\equiv{1\over\sqrt{2\pi}\Delta\tau}\int_{-\infty}^\infty \cosh\eta d\eta\int_{-\infty}^\infty\tau d\tau 
\exp( - {\eta^2 \over 2 (\Delta \eta)^2})
 \exp (- {(\tau-\tau_0)^2 \over 2(\Delta \tau)^2})\exp(-\gamma_n(b)\cosh\eta)
(q_0\beta)^2\tau^2\cosh^2\eta\non
&&
\approx(3\tau_0\Delta\tau^2+\tau_0^3)(q_0\beta)^2\exp({1\over\Delta\eta^2})f_3(\xi_n), 
f_3(\xi_n)=2{K_2(\xi_n)\over \xi_n}=
2\left(K_0(\xi_n)/\xi_n+K_1(\xi_n)(1+{2\over \xi_n^2})\right),\\
&& I_1\equiv{1\over\sqrt{2\pi}\Delta\tau}\int_{-\infty}^\infty d\eta\cosh\eta\int_{-\infty}^\infty\tau d\tau 
\exp( - {\eta^2 \over 2 (\Delta \eta)^2})
 \exp (- {(\tau-\tau_0)^2 \over 2(\Delta \tau)^2})\exp(-\gamma_n(b)\cosh\eta)
\tau^2\sinh^2\eta\non&&
\approx \tau_0(3\Delta \tau^2+\tau_0^2)\exp({1\over\Delta\eta^2})f_1(\xi_n), 
f_1(\xi_n)=2(K_2(\xi_n)/\xi_n)=2K_0(\xi_n)/\xi_n+4K_1(\xi_n)/\xi_n^2.
\eea
Then \bea &&\int\;d^4xS(x,K,q)=\tau_0\exp({1\over\Delta\eta^2})\times\non&&
\left[\Phi_{12}-iq_o\beta(\tau_0+\Delta\tau^2/\tau_0)F_2(K_T)-{1\over2}q_o^2\beta^2
(3\Delta\tau^2+\tau_0^2)F_3(K_T)-q_L^2/2F_1(K_T)\right],\label{bigsint}\eea
where \bea &&\Phi_{12}=\sum_{n=1}^\infty\int\;d^2bf_0(\xi_n)
B_n(\bfb,\bfK_T)\psi^{(-)}_{{\bfp}_1}(\bfb)\psi^{(-)*}_{{\bfp}_2}(\bfb)\label{phi12int}\\
&&F_{2}(K_T)=\sum_{n=1}^\infty\int\;d^2bf_2(\xi_n)
B_n(\bfb,\bfK_T)\vert\psi^{(-)}_{K_T}(\bfb)\vert^2\\
&&F_{3}(K_T)=\sum_{n=1}^\infty\int\;d^2bf_3(\xi_n)
B_n(\bfb,\bfK_T)\vert\psi^{(-)}_{K_T}(\bfb)\vert^2\\&&
F_1(K_T)=\sum_{n=1}^\infty\int\;d^2bf_1(\xi_n)B_n(\bfb,\bfK_T)\vert\psi^{(-)}_{K_T}(\bfb)\vert^2\eea
The denominator is obtained by evaluating the emission function (\ref{sdenom}) and the following functions enter:
\bea&&F_0(K_T)=\sum_{n=1}^\infty\int\;d^2bf_0(\xi_n)B_n(\bfb,\bfK_T)\vert\psi^{(-)}_{K_T}(\bfb)\vert^2\label{f0def}\\
&&\Phi_{ii}=\sum_{n=1}^\infty\int\;d^2bf_0(\xi_n)
B_n(\bfb,\bfK_T)\vert\psi^{(-)}_{{\bfp}_i}(\bfb)\vert^2.\label{phiiint}\eea
With these definitions, one may  show:
\bea
&&C(\bfq+q_L{\bf\hat{z}},\bfK_T)=1-q_l^2R_l^2-q_o^2\beta^2{\widetilde{\Delta\tau}}^2
+{\vert\Phi_{12}\vert^2\over\Phi_{11}\Phi_{22}}\label{cdef}\\
&&R_l^2=(3\Delta\tau^2+\tau_0^2)F_1(K_T)/F_0(K_T)\\&&
{\widetilde{\Delta\tau}}^2=(3\Delta\tau^2+\tau_0^2)F_3(K_T)/F_0(K_T)-
(\tau_0+\Delta\tau^2/\tau_0)^2
\left\vert{F_2(K_T)\over F_0(K_T)}\right\vert^2.\eea

The spectra are given by Eq.~(\ref{spectra0}) \cite{Wiedemann:1999qn}
\bea
E_p{dN\over d^3p}={dN\over dY M_\perp dM_\perp d\phi_p}=\int d^4x S(x,p),\eea
so that in general
\bea
{dN\over   dM^2_\perp}={1\over2} (2\pi)) \int dY\int d^4x S(x,p),\label{spectra}\eea
in which the azimuthal symmetry of the angular distribution is used.

The STAR detector receives pions 
for values of $Y$ between $\pm0.5$ and presents its results
in terms of $\langle{dN\over 2\pi M_\perp  dM_\perp dY}\rangle_{\vert Y\vert<0.5}$ 
which is given by 
\bea\left\langle{dN\over 2\pi M_\perp dM_\perp dY}\right\rangle=\int_{-0.5}^{0.5}dY\int\;d^4x S(x,K).\eea
Numerical studies showed us that the average  over $Y$ is extremely well approximated
by simply replacing $Y$ by 0. This requires $\delta\eta>1/2,$ and we use
$\delta\eta\approx 1$. Thus we find 
\bea\left \langle{dN\over 2\pi M_\perp  dM_\perp dY}\right \rangle_{\vert Y\vert<0.5}\cong{\tau_0\over 8\pi^3}\exp({1\over\Delta\eta^2})F_0(\bfK_T).\eea

The evaluation proceeds by reducing the two dimensional integrals
of  Eqs.~(\ref{phi12int}-\ref{phiiint}) to those of one dimension by an analytic evaluation of 
 the angular
integrals. We need the partial wave expansions
\bea
{\psi_{\bfp_1}^{(-)}}^*
({\bfb})=f_0(p_1,b)+2\sum_{m=1,\infty}f_m(p_1,b){(-i)}^m\;\cos{m\phi_1}
\label{psim1}\\
\psi_{\bfp_2}^{(-)}
({\bfb})=f_0^*(p_2,b)+2\sum_{m=1,\infty}{f_m}^*(p_2,b){(i)}^m\;\cos{m\phi_2}
,\label{psim2}\eea
where  $\cos\phi_i=\widehat{\bfp}_i\cdot\widehat{\bfb}.$

We encounter integrals of the form
\bea
A_{mn}(z)=\int_0^{2\pi}d\phi\;e^{z\cos\phi}\cos m\phi_1\cos n\phi_2
.\label{12}\eea
For $\bfq\parallel \bfK,\;\bfq=\bfq_o,\;\phi_1=\phi_2=\phi$, so that we define
\bea
&&A_{mn}(\parallel,z)\equiv\int_0^{2\pi}d\phi\;e^{z\cos\phi}\cos m\phi
\cos n\phi\\
&&=\pi \left( I_{m+n}(z)+I_{m-n}(z)\right),\label{para}
\eea 
where $I_{m\pm n}(z)$ are modified Bessel functions:
\bea
I_n(z)=(-i)^n J_n(i z),
\eea for real $z$, and  $I_n(z)=I_{-n}(z)$.
An  integral representation is
\bea I_n(z)={1\over2\pi}\int_0^{2\pi} d\phi e^{z\cos\phi} \cos{n\phi}.\eea
For $\bfq\cdot\bfK=0$, $\bfq=\bfq_s$, we define
\bea \cos\alpha={K\over\sqrt{K^2+q^2/4}},\qquad\sin\alpha={q/2\over
\sqrt{K^2+q^2/4}}.\eea
Then
\bea
&&A_{mn}(\perp,z)=\pi \left( I_{m+n}(z)\cos{(m-n)\alpha}+
I_{m-n}(z )\cos{(m+n)\alpha}\right).\label{perp}
\eea
Note that for the denominators, one gets integrals in which
the two angles of  (\ref{12}) are the same. Then the expression (\ref{para}) is to be used.

The use of (\ref{para}) and (\ref{perp}) in (\ref{12int}) yields
\bea
\Phi_{12}(\parallel,\perp)
&=&M_\perp(K)\int_0^\infty bdb\rho(b)e^{-M_\perp(K)\cosh\eta_t(b)/ T}\nonumber\\
&&\times\left[\sum_{m,n=0}^\infty\epsilon_m\epsilon_n f_m(p_1,b)
f_n^*(p_2,b)(i)^{n-m}A_{mn}((\parallel,\perp), {K\over T}\sinh{\eta_t(b)})\right]\label{firstphi}\\
\epsilon_0&=&1,\quad\epsilon_{n>0}=2.\eea The notation $(\parallel,\perp)$ 
denotes either of 
the two possibilities $\bfq\parallel\bfK,\bfq\perp\bfK$.
Similarly,
\bea
&&\Phi_{ii}(\parallel,\perp)
=M_\perp(p_i)\int_0^\infty bdb\rho(b)e^{-M_\perp(p_i)\cosh\eta_t(b)/T}\nonumber\\
&&\times\left[\sum_{m,n=0}^\infty\epsilon_m\epsilon_n f_m(p_i,b)f_n^*(p_i,b)(i)^{n-m}
A_{mn}((\parallel,\perp), K\sinh{\eta_t(b)}/T)\right]\label{lastphi}
\eea

\subsection {Exact numerical evaluation of correlation function}
\label{subs:ee}
We present an alternate calculational technique that gives the correlation function
for any value of $\bfq$. To do this, start by recalling Eqs.~(\ref{cli}) 
and (\ref{sdefy0}) and note the appearance of 
 the  integral
\bea
&&I_{\tau\eta}\equiv\non
&& {1\over\sqrt{2\pi}\Delta\tau}\int_{-\infty}^\infty 
d\eta\cosh\eta \exp(-\gamma_n(b)\cosh\eta)\exp( - {\eta^2 \over 2 (\Delta \eta)^2})
\non&&\int_{-\infty}^\infty\tau d\tau 
\exp (- {(\tau-\tau_0)^2 \over 2(\Delta \tau)^2})
e^{i\tau\left((\omega_2-\omega_1)\cosh \eta-iq_L\sinh\eta\right)}\non 
&& =\int_{-\infty}^\infty 
d\eta\cosh\eta \exp(-\gamma_n(b)\cosh\eta+i\alpha\tau_0)\exp(-{\eta^2 \over 2 (\Delta \eta)^2})
(\tau_0 +i\alpha\Delta\tau^2)\exp(-\alpha^2\Delta\tau^2/2),\non\label{resultau}\\
&&\alpha\equiv(\omega_2-\omega_1)\cosh \eta-q_L\sinh\eta.
\eea The procedure of this section is to evaluate 
the   integral over $\eta$  numerically. 

Let's set up the full calculation. Integrating over $\tau$ and using the result (\ref{resultau})
yields
\bea &&\int\;d^4x S(x,K,q)=\sum_{n=1}^\infty\int d^2bB_n(\bfb,\bfK_T))\psi_{\bfp_1}^{(-)}(\bfb)
\psi_{\bfp_2}^{(-)*}(\bfb)
\times\non&&
\int_{-\infty}^\infty
d\eta\cosh\eta 
\exp(-\gamma_n(b)\cosh\eta+i\alpha\tau_0)\exp( - {\eta^2 \over 2 (\Delta \eta)^2})
(\tau_0 +i\alpha\Delta\tau^2)\exp(-\alpha^2\Delta\tau^2/2),\non&&
\eea

There are some remarks to be made here: there is no need to make the approximation
of Eq.~(\ref{fap}), and  
the present  expression is actually much more compact than (\ref{bigsint}).  
Possible cross terms \cite{Chapman:1994yv} involving $q_Oq_L$ are small for the experiments
with $Y\approx0$ that we analyze, so 
we neglect the cross terms and take either $q_L=0$ or $q_O=0$. 
Then the integrands
have terms either even or odd in $\eta$. The odd terms cancel. 
So  define (with $\Delta\omega\equiv\omega_2-\omega_1)$)

\bea 
&&I(\gamma_n(b),\Delta\omega,q_L,\Delta\eta,\Delta\tau,\tau_0)\equiv\non&&
\int_{-\infty}^\infty
d\eta\cosh\eta 
\exp(-\gamma_n(b)\cosh\eta+i\alpha\tau_0)\exp( - {\eta^2 \over 2 (\Delta \eta)^2})
(\tau_0 +i\alpha\Delta\tau^2)\exp(-\alpha^2\Delta\tau^2/2).\eea
We use various  specific values of the arguments of the function $I$
 to compute the different observables. According to Eqs.~(\ref{logform},\ref{quadform})
a radius $R_i$ can be computed using $q_i\ne 0,q_{j\ne i}=0$. Thus to compute $R_O$ we take
$q_{L,S}=0$, so that 
\bea
&&I(\gamma,\Delta\omega,0,\Delta\eta,\Delta\tau,\tau_0)\equiv I_O(\gamma,\Delta\omega,\Delta\eta,\Delta\tau,\tau_0)\equiv\non&&
 2\int^\infty_0 d\eta\cosh\eta 
\exp(-\gamma\cosh\eta)\exp( - {\eta^2 \over 2 (\Delta \eta)^2})
e^{iz\tau_0}(\tau_0+iz\Delta\tau^2)\exp(-z^2\Delta\tau^2/2)\non
&&z\equiv \Delta\omega\cosh\eta.\eea
To compute $R_L$ we take $\Delta\omega=0$, so that
\bea&&I(\gamma,0,q_L,\Delta\eta,\Delta\tau,\tau_0)\equiv I_L(\gamma,q_L,\Delta\eta,\Delta\tau,\tau_0)\equiv\non&&
 2\int^\infty_0 d\eta\cosh\eta 
\exp(-\gamma\cosh\eta)\exp( - {\eta^2 \over 2 (\Delta \eta)^2})
(\tau_0\cos(y\tau_0)-y\Delta\tau^2\sin(y\tau_0))
\exp(-y^2\Delta\tau^2/2)\non
&&y\equiv -q_L\sinh\eta
\eea
In  computing $R_S$, the spectra, or the denominator of the correlation function   we have $\Delta \omega=0,q_L=0$. Then  we use:
\bea
&&I(\gamma,0,0,\Delta\eta,\Delta\tau,\tau_0)\equiv
 2\tau_0\int^\infty_0 d\eta\cosh\eta 
\exp(-\gamma\cosh\eta)\exp( - {\eta^2 \over 2 (\Delta \eta)^2})
.\label{icode}\eea
Then the correlation function is given by
\bea &&C(\bfq+q_L{\bf\hat{z}},\bfK)=1+{\left\vert \chi_{12}\right\vert^2\over\chi_{11}\chi_{22}}\\
&&\chi_{12}=\int d^4xS(x,K,q)\non&&=
\sum_{n=1}^\infty\int d^2bB_n(\bfb,\bfK_T)I(\gamma_n(b),\Delta\omega ,q_L,\Delta\eta,\Delta\tau,
\tau_0)\psi_{\bfp_1}^{(-)}(\bfb)
\psi_{\bfp_2}^{(-)*}(\bfb)\\&&
\chi_{ii}=
\sum_{n=1}^\infty\int d^2bB_n(\bfb,\bfp_i)I(\gamma_n(b),0,0,\Delta\eta,\Delta\tau
,\tau_0)\left\vert\psi^{(-)}_{{\bfp}_i}(\bfb)\right\vert^2\eea
 For very small values of
$\bfq$ this correlation function reduces to the one obtained in second order, (\ref{cdef}).
There are two changes. If $\bfq$ is not small then one obtains the correct correlation function. A 
second
change is that the approximation (\ref{fap}) is not used. The 
differences between using the procedure of
this subsection and that 
of the previous subsection are negligible ($<2\%$ and indistinguishable)   for the cases
 we have studied. This provides a further verification of the approximations used 
in Sect.~(\ref{subs:expeval}),
 but the present formalism avoids those approximations.

\section{Applications}
\label{sec:apps}

The description of the formalism is essentially complete.
The plane wave emission function $S_0$
is defined in Sect.~II and its symmetry discussed in Sect. IV. The distorted 
wave emission function function $S$ is defined in Sect.~III, and its evaluation
elaborated in Sect.~VII. The optical potential is defined in Sec.~V, and its
use in a wave equation to obtain the distorted wave is discussed in Sec.~VI.
The correlation function is evaluated in Sect. VIII.


Our technique  
 may be compared
with the Buda-Lund model,   an 
efficient representation of the data\cite{Buda}, in which
 the temperature and 
fugacity are taken as position-dependent functions appearing
in a Boltzmann distribution. The effects of our optical potential could provide
an explanation of some of those deduced dependencies. 

We are now ready to confront the data.
\subsection {DWEF Fits to Central Au+Au Collisions}
\label{subs:num}

\small
\begin{table*}
  \caption{Best fit parameters (F193) used in the calculations shown, with variances; fits use Eqns.~(\ref{ckqgauss1}, \ref{logform}).  {\bf Bold-face} indicates parameters not varied in the fitting procedure (see text).}
  \vspace{0.1cm}
  \begin{tabular}{|lllllllllll|}\hline
   $~~~~T$ & $~~~\eta_f$ & $~~~\Delta\tau$ &~~$R_{WS}$ & $~a_{WS}$& $~~~w_0$& $~~~~~~w_2$ &~~~$\tau_{0}$ &~~~~$\Delta\eta$ &~~$~~\epsilon$ & $~~~~\mu_\pi$\\
   $(MeV)$ & & $(fm/c)$ &~~$(fm)$ & $(fm)$& $(fm^{-2})$& & $(fm/c)$ & & & $(MeV)$\\
\hline
{\bf 193.00}   &~~1.507 &~~~2.382 &~11.773 &~~0.909 &~~0.139 &~~0.847~+$i$0.120 &~~~8.60 &~~~~0.991&~~{\bf 0.000}&~~{\bf 139.57}\\
              & $\pm$0.025 &~~$\pm$0.07 &~~$\pm$0.06 & $\pm$0.015 & $\pm$0.046 & $~\pm$0.014~$\pm$0.002 & $~~\pm$0.10 &~~~$\pm$0.032& & \\
     \hline
      \end{tabular}
      \end{table*}
\normalsize

Table 1 gives the parameters of our best fit to the STAR data for Au+Au central collisions at 200 GeV, which we will refer to as ``F193''.
For this fit to the data, as shown in Figs. \ref{one} and \ref{three}, the $\chi^{2}$ is 56.45, and the $\chi^{2}$ per degree of freedom is 2.45.
Table I also gives the estimated variances of those fit parameters that were varied, as calculated by determining the parameter variation required to increase the $\chi^{2}$ value by one unit.  Correlations between different parameters are not considered, but, as will be discussed below, more than one set of parameters can produce a quality fit to the data.

\begin{figure}
\includegraphics[width=9 cm] {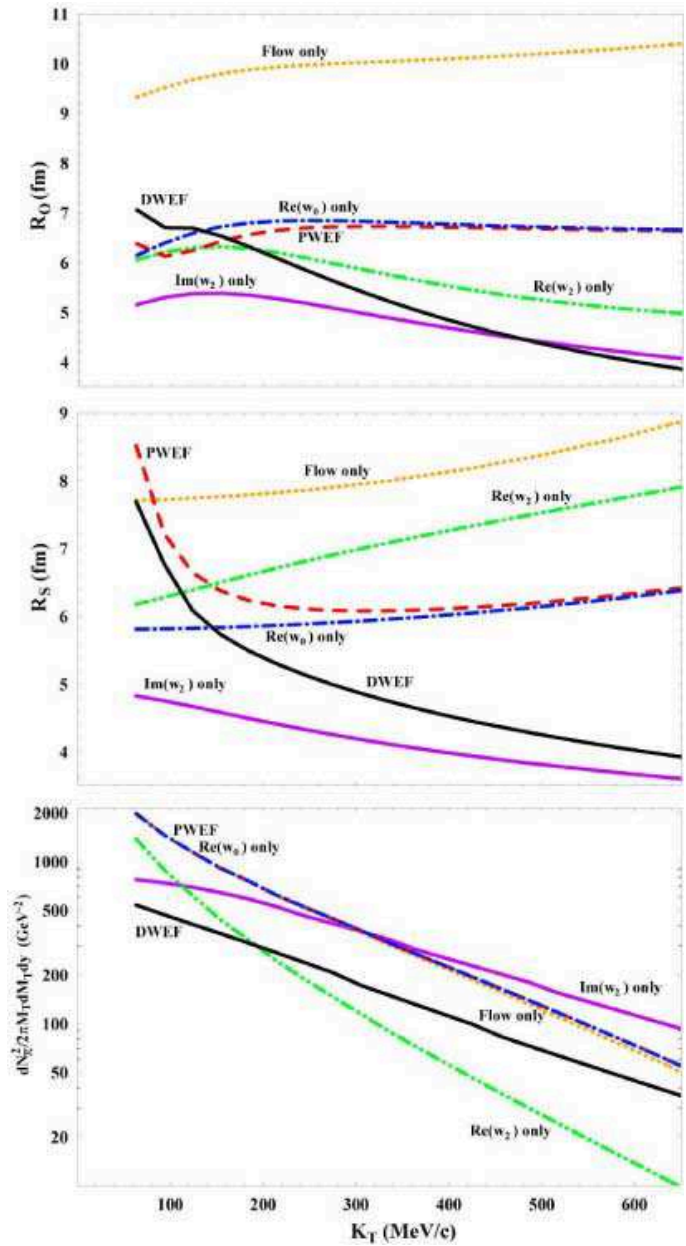}
\caption{\label{rout} (Color online) Calculations of $R_O$, $R_S$, and spectrum, isolating potential and 
flow effects.  DWEF (black solid) full calculation; PWEF (red dashed) 
plane wave calculation with no optical potential or flow; $Re(w_2)$ only (green dot-dot-dash) 
calculation with no flow and only real part of momentum dependent optical potential ($w_2$); 
$Re(w_0)$ only (blue dot-dash) calculation with no flow and only real constant optical potential ($w_0$);  
$Im(w_2)$ only (solid violet) calculation with no flow and only imaginary part of momentum-dependent optical 
potential ($w_2$); 
Flow only (orange dotted) 
calculation with flow but no optical potential.  
Non-zero parameters in all calculations have the F193 values of Table I.} 
\end{figure}

To illustrate the importance of the various effects of the DWEF model for computing
the radii $R_O, R_S$ and the spectrum, we have done calculations using the F193 fit parameters with various effects switched on separately.  This is shown in Fig.~\ref{rout}.  The curves labeled DWEF show the full calculation.  Those labeled PWEF are computed using plane waves, i.e., the optical potential ($w_{0,2}$) and flow ($\eta_{F}$) are set to zero.   The curves labeled $Re(w_{0})$ and $Re(w_{2})$ use only the real constant or momentum-dependent parts of the optical potential, respectively, set to the F193 values, with no flow.   The curves labeled $Im(w_{2})$ use only the imaginary momentum-dependent part of the optical potential set to the F193 value, with no flow.   Finally, the curves labeled $flow$ uses the F193 value of $\eta_{F}$, with no optical potential.  All other parameters in these calculations are set to the F193 values of Table 1.  This study indicates that both flow and the optical potential modify the HBT radii, but only the momentum dependent parts of the optical potential ($w_2$) affect the spectrum.

Figs.~\ref{one} and \ref{three} show DWEF calculations using F193, as compared with the STAR data \cite{STARHBT,STARspec} used in the fit.  We see that both the HBT radii and the spectrum (including the spectrum normalization) are
reproduce very well indeed.  However, some clarification is needed here on the issue of resonance pions.  Our model describes {\it only} those ``direct'' pions that are emitted directly within the fireball and that participate fully in the HBT correlation.  It does {\it not} predict the ``halo'' fraction of pions originating from resonance decays later in the process that will not participate in the HBT correlation (or that would make an unmeasurable ``spike'' in the correlation function near $q=0$).  These latter pions are present in the measured spectrum and must be removed before fitting the spectrum with a DWEF calculation.

\begin{figure}
\includegraphics[width=17cm]   {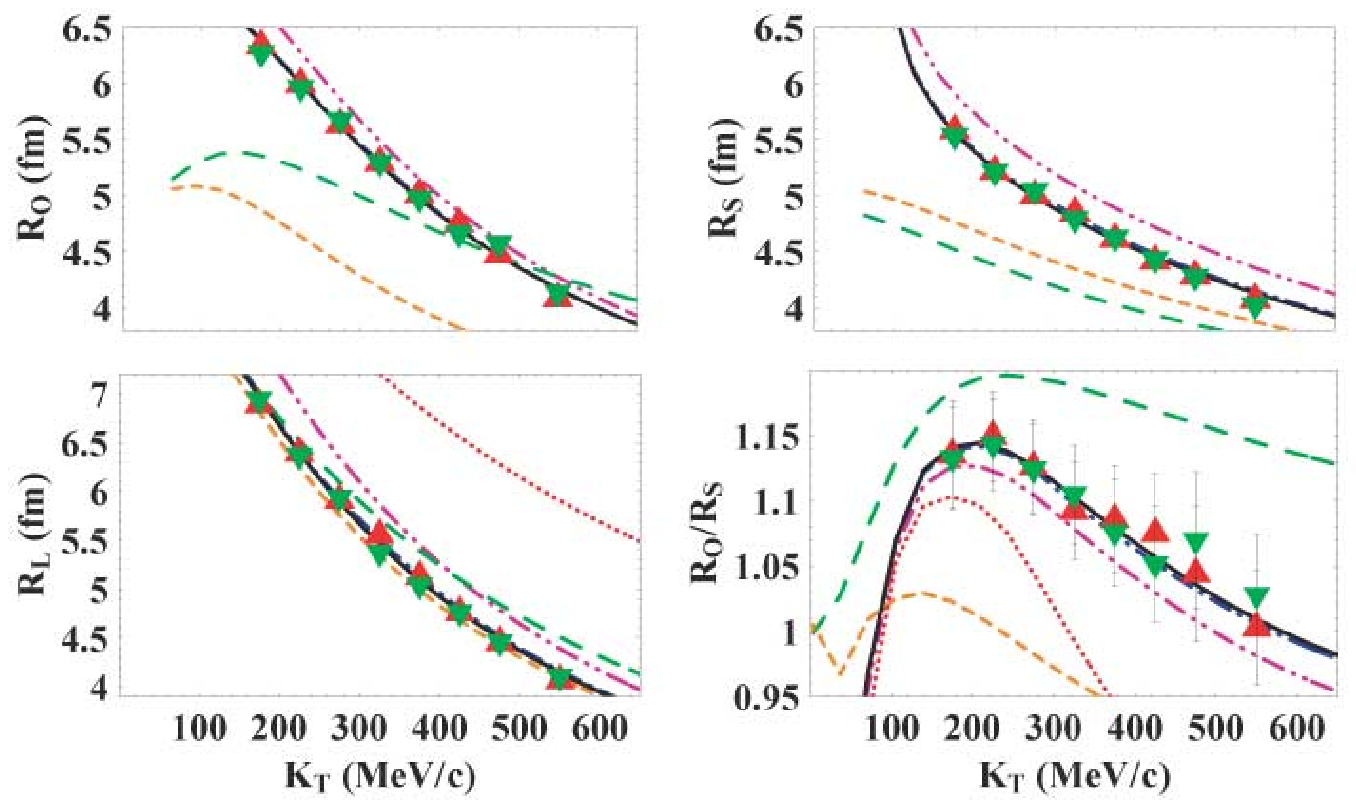}
\caption{\label{one}  (Color online) HBT Radii $R_s, R_o,$, $R_l$ and the ratio $R_o/R_s$; 
Data \cite{STARHBT}): $\nabla$ (green) $\Rightarrow \pi^{+}\pi^{+}$; 
$\triangle$ (red) $\Rightarrow \pi^{-}\pi^{-}$.  
Curves: solid (black) $\Rightarrow$ full calculation; 
dotted (red) $\Rightarrow\eta_f=0$ (no flow); dashed (orange) $\Rightarrow$ Re$[U]$=0 (no refraction); 
dashed (green) $\Rightarrow$ 
$U$=0 (no potential); dot-dashed (blue)$\rightarrow,\;\mu_\pi=0$ (almost same as solid)
 double-dot-dashed (violet) $\Rightarrow$ substituting Boltzmann for Bose-Einstein thermal distribution.}
\end{figure}

\begin{figure}
\includegraphics[width=15cm]  {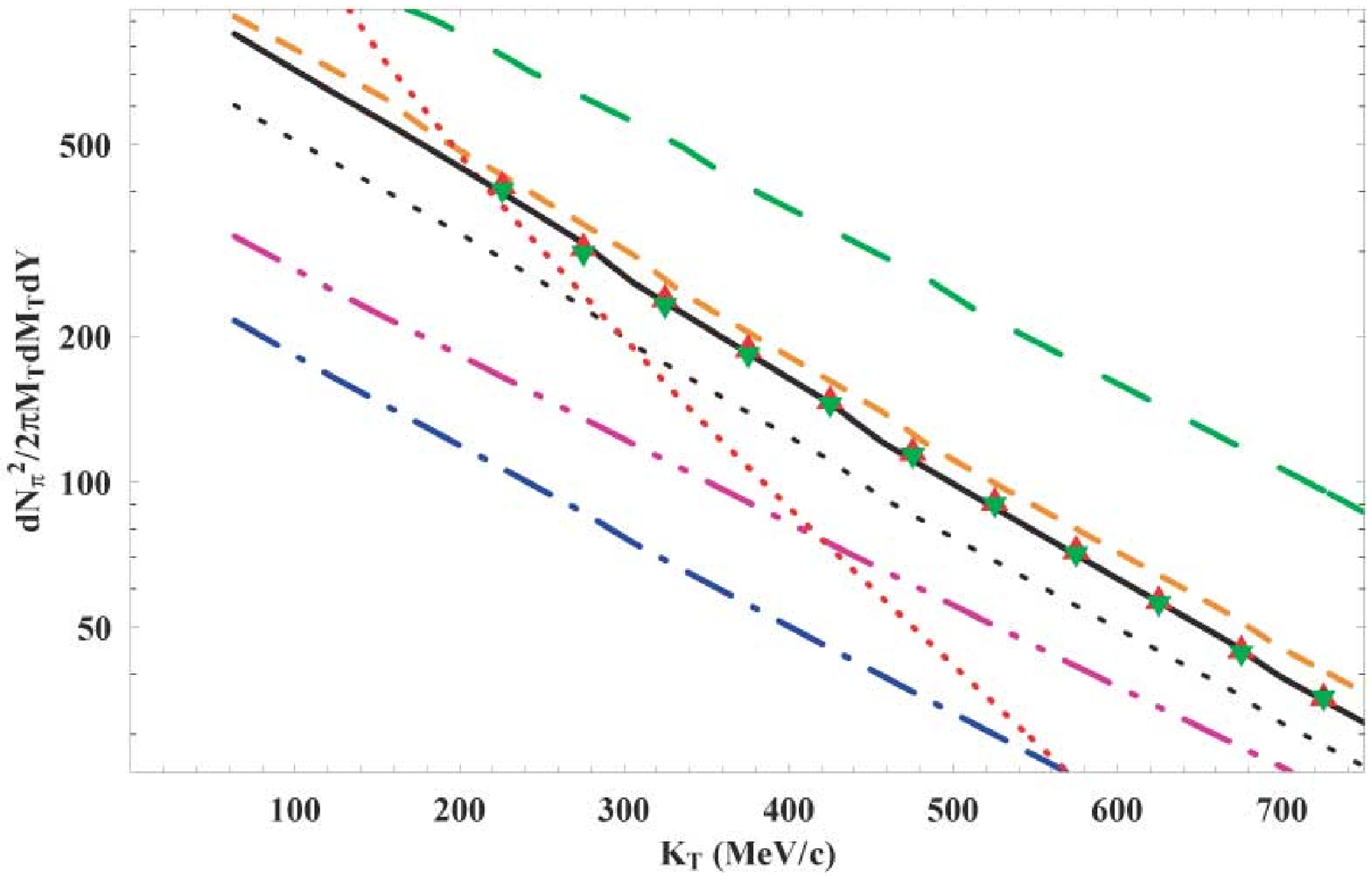}
\caption{\label{three} (Color online) 
Pion momentum spectrum.
Data \cite{STARspec}: $\nabla$ (green) $\Rightarrow \pi^{+}$; $\triangle$ (red) $\Rightarrow \pi^{-}$. 
Black dotted line is the `non-resonant pion'' spectrum used in the fit. The curves are denoted
as in Fig.~\ref{one}.                                                                                                       
}
\end{figure}

We have done this by using the HBT $\lambda$ parameter previously
 extracted as part of the HBT analysis, which we take to represent 
the probability that both correlated pions are direct rather than halo pions. 
 We fit the extracted $\lambda$ parameter with a straight line in transverse
 momentum to obtain the function $\lambda(p_T)$.  Then we remove the 12\% weak decay correction of the published pion spectrum\cite{STARspec} and multiply it by $\sqrt{\lambda(p_T)}$, thereby producing an estimate of the momentum spectrum of direct pions only.  This corrected spectrum, as indicated by the black small-dashed line in Fig. \ref{three}, is then used in the data fitting.  Following the DWEF calculation, the predicted theoretical spectrum is ``uncorrected" by dividing it by $\sqrt{\lambda(p_T)}$ and then reducing it by 12\%, so that it can be compared directly with the published data.  These predictions, now including halo pions, are the curves shown in the spectrum plots.  We note that the effects of the two corrections to the spectrum tend cancel each other, and that their net effect on the fits is primarily an increase in temperature and flow parameters.  We note also that the Phobos Collaboration has published spectrum data in the low $p_T$ region\cite{Back:2004uh}, but because the value of $\lambda$ in this region are not known, we cannot make similar corrections and therefore have not included these data in our current analysis.  The DWEF fit parameters, particularly the temperature, flow, chemical potential, and Woods-Saxon diffuseness, are sensitive to the details of our treatment of resonances, thus introducing a procedure-dependent systematic error in DWEF parameter extraction.

Our DWEF calculations predict the correlation function of two pions with a particular vector momentum difference $\bfq$.  There are at least two ways of extracting HBT radii from such correlation functions. One is to use the quadratic form (\ref{ckqquad}) of the correlation function at very small magnitudes of $\bfq$.  The other is to use the Gaussian form (\ref{ckqgauss1}) at larger values of $\bfq$ that correspond to the falloff region of the correlation function.
In our previous publication\cite{Cramer:2004ih} we used the former  method with $q=K_T/40$, a technique which essentially extracts HBT radii from the calculated curvature of the correlation function near $q=0$. In the present work we use the Gaussian form with $q=0.15$ fm$^{-1}$, a momentum difference that evaluates the correlation function near its half-maximum point, a procedure that bears more resemblance to the extraction of radii from experimental data with Gaussian fitting.  We have found that in the medium and high momentum regions the two procedures gives very similar descriptions of the data and extract similar radii.  However, we have found that for values of $K_T < 175$ MeV/c, the two methods diverge, and that the Gaussian form gives more reliable radii.

\subsection {Wave Function Plots}
\label{subs:wf}

\begin{figure}
\includegraphics[width=15.5 cm]{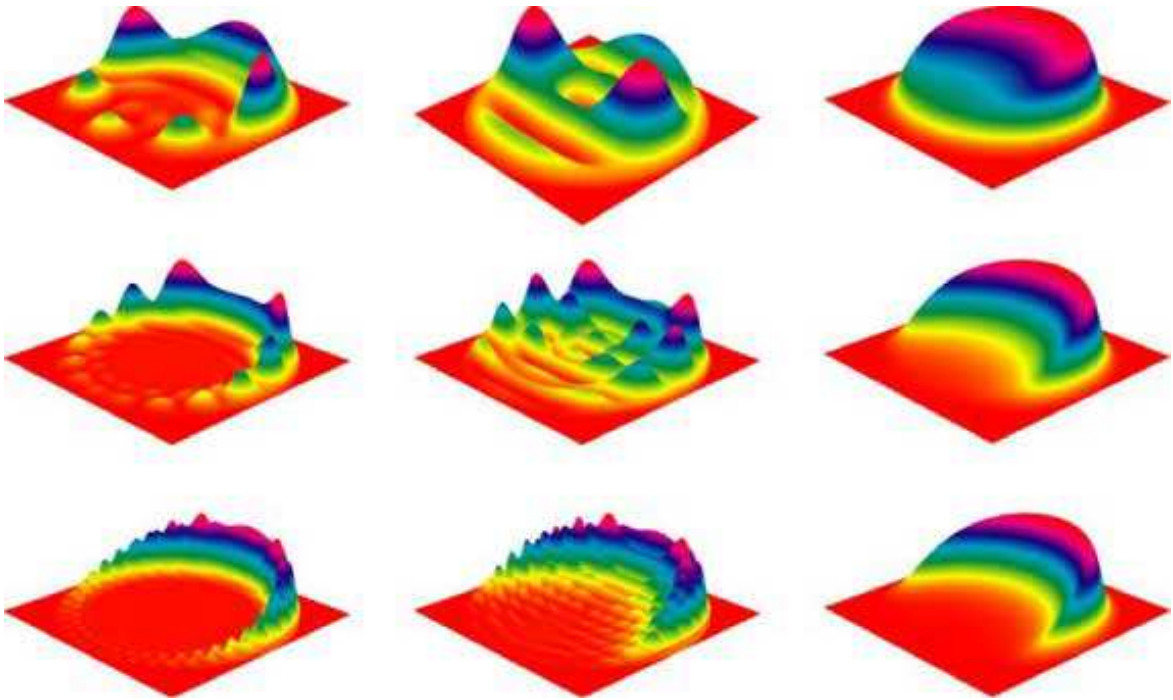}
\caption{\label{wf}  (Color online) Wave functions. The figures show the absolute square of the
calculated wave functions times the density $\rho(b)$. For each (row) 
value of $K_T~(100,250,600)$ MeV/c, the full
calculation, the calculation including only the imaginary part of the optical potential and
the eikonal approximation with the same absorption are  shown horizontally. The out direction is parallel to the lower right
edge of each picture.}
\end{figure}

We can understand more about how HBT measurements in various momentum regions probe the system under study by examining the computed wave functions from the DWEF calculation.  Such wave functions calculated with the F193 fit are shown in Fig.~\ref{wf}.
The complete DWEF wave functions (left column) differ significantly from the semi-classical eikonal approximation calculated with the same absorption (right column) at each
of the energies we consider. The importance of the real part of the optical potential is illustrated by comparison with the center column, showing wave functions calculated with the real part of the optical potential set to zero.  The maxima and minima of these plots result from interference effects caused by the optical potential as incorporated by solving the quantum mechanical wave equation.  Their spacing gives a qualitative indication of the pion wavelength in the medium. The differences between the quantum calculations and the semi-classical eikonal model grow smaller as the value of $K_T$ increases. 
We note that while the high momentum wave functions sample only a ``bright ring'' 
at the edge of the fireball, the low momentum wave functions is also non-vanishing  in  
most parts of the fireball.

\subsection {Temperature Ambiguities}
\label{subs:tempamb}

In our previous publication\cite{Cramer:2004ih} the temperature
 parameter used in the pion emission function was $T=222$ MeV. 
 This is an uncomfortably large value for the temperature of the
 medium, high enough that pions would be expected to ``melt" in 
such an environment and lose their identity.  Therefore, we 
 explored (not shown)  the extent to which the DWEF model {\em requires} such large 
values of temperature and   
 showed  that there is a 
parameter ambiguity involving the parameters $T$ and $a_{WS}$, 
such that a fit of reasonable quality could be obtained for temperature 
values over a fairly broad range.

\small
\begin{table*}
  \caption{Nine Fit-Parameters Sets: 
These fits were obtained by fixing a range of values for the temperature $T$ and fitting the central Au+Au 200 GeV STAR data.  {\bf Bold-face} indicates a parameter fixed for the search.}
  \vspace{0.1cm}
  \begin{tabular}{|lllllllllllll|}\hline
~Fit~~&~~$~~~~T$ &~~~$\eta_f$ &~~~$\Delta\tau$ & $~R_{WS}$ & $~a_{WS}$& $~~~w_0$& $~~~~~~w_2$
&~~~~~$\tau_{0}$ &~~$\Delta\eta$ &~~$~~~~\epsilon$ &~~~~~$\mu_\pi$&~~$\chi^{2}/DOF$\\
&~~$(MeV)$ & & $(fm/c)$ & $~(fm)$ & $~(fm)$& $(fm^{-2})$& &~~$(fm/c)$ &~~ &~~ &~~~$(MeV)$&\\
\hline
F173  &~~{\bf 173.0}   &~1.451 &~~2.664 & 11.924 &~~0.541 &~0.095 & 0.777~+$i$0.112 &~~~~8.77 &  0.983  &~~~{\bf 0.000} &~~~139.57 &~~~~~2.81\\
F183  &~~{\bf 183.0}   &~1.475 &~~2.479 & 11.811 &~~0.754 &~0.123 & 0.815~+$i$0.113 &~~~~8.66 &  0.994  &~~~{\bf 0.000} &~~~139.51 &~~~~~2.64\\
F193  &~~{\bf 193.0}   &~1.507 &~~2.382 & 11.773 &~~0.909 &~0.139 & 0.847~+$i$0.120 &~~~~8.60 &  0.991  &~~~{\bf 0.000} &~~~{\bf 139.57} &~~~~~2.45\\
F193a &~~{\bf 193.0}   &~1.508 &~~2.380 & 11.773 &~~0.912 &~0.139 & 0.850~+$i$0.121 &~~~~8.62 &  0.987  &~~~0.0002 &~~~{\bf 139.57} &~~~~~2.56\\
F193b &~~{\bf 193.0}   &~1.510 &~~2.378 & 11.770 &~~0.939 &~0.140 & 0.854~+$i$0.120 &~~~~8.67 &  0.977  &~~~0.0137 &~~~{\bf 139.57} &~~~~~2.56\\
F203  &~~{\bf 203.0}   &~1.565 &~~2.200 & 11.935 &~~1.046 &~0.127 & 0.823~+$i$0.115 &~~~~8.66 &  0.990  &~~~{\bf 0.000} &~~~129.61 &~~~~~3.45\\
F203a &~~{\bf 203.0}   &~1.536 &~~2.330 & 11.641 &~~1.271 &~0.146 & 0.825~+$i$0.100 &~~~~8.62 &  0.967  &~~~0.263 &~~~134.44 &~~~~~2.84\\
F213  &~~{\bf 213.0}   &~1.577 &~~2.323 & 11.689 &~~1.330 &~0.131 & 0.817~+$i$0.101 &~~~~8.54 &  0.970  &~~~0.296 &~~~129.54 &~~~~~3.16\\
F220  &~~{\bf 220.0}   &~1.616 &~~2.328 & 11.802 &~~1.355 &~0.115 & 0.795~+$i$0.103 &~~~~8.47 &  0.977  &~~~0.297 &~~~124.88 &~~~~~4.52\\
     \hline
      \end{tabular}
      \end{table*}
\normalsize

Therefore, it is desirable to choose a temperature appropriate 
to the medium.  If one takes quite literally the expectation that 
the DWEF model describes the {\it initial} emission of pions and that the first pions are 
produced directly in the strongly interacting quark-gluon plasma as it make a phase 
transition to a hadronic phase, then the emission function should have the temperature 
of the QGP transition.  Similarly, if the pions are produced as massless objects due to 
chiral symmetry restoration in the medium, then the chemical potential should equal the free
 pion mass.  Recent lattice gauge calculations reported at Quark Matter 2005\cite{Katz:2005br}
 give the critical QGP transition temperature to be 193 MeV.  Therefore, we adopt (Table I)  T=193 MeV 
and $\mu_{\pi}$=139.57 MeV and search for a new fit to the STAR $\sqrt{S_{NN}}$=200 GeV
Au+Au data.  To our surprise, the fit we obtain (see Figs. 3 and 4) with parameters fixed at these values is the best we have found, with an overall $\chi^{2}$ of 56.45 and a $\chi^{2}$ per degree of freedom 
of 2.45.  Further, subsequent searches in which the fixed temperature was set to other values 
between 173 MeV and 220 MeV show that there is a definite minimum in $\chi^{2}$ at just the 
value of temperature given by the lattice gauge calculation.  Table II shows nine different 
set of fit parameters, all of which give good fits to the central STAR $\sqrt{S_{NN}}$=200 
GeV Au+Au data.  In Table II, the fixed parameters  are 
indicated in bold face.  We see that F193, the parameter set of Table I, gives the best fit, 
and that for temperatures greater than 193 MeV it is necessary to use a non-zero value of 
$\epsilon$, invoking a Kisslinger-type wave equation \eq{u526}, to obtain a quality fit.  
However, fits F193a and F193b indicate that searching on $\epsilon$ as part of a search 
at $T =$ 193 MeV gives values near zero and did not improve the fit.

\begin{figure}
\includegraphics[width=10cm]  {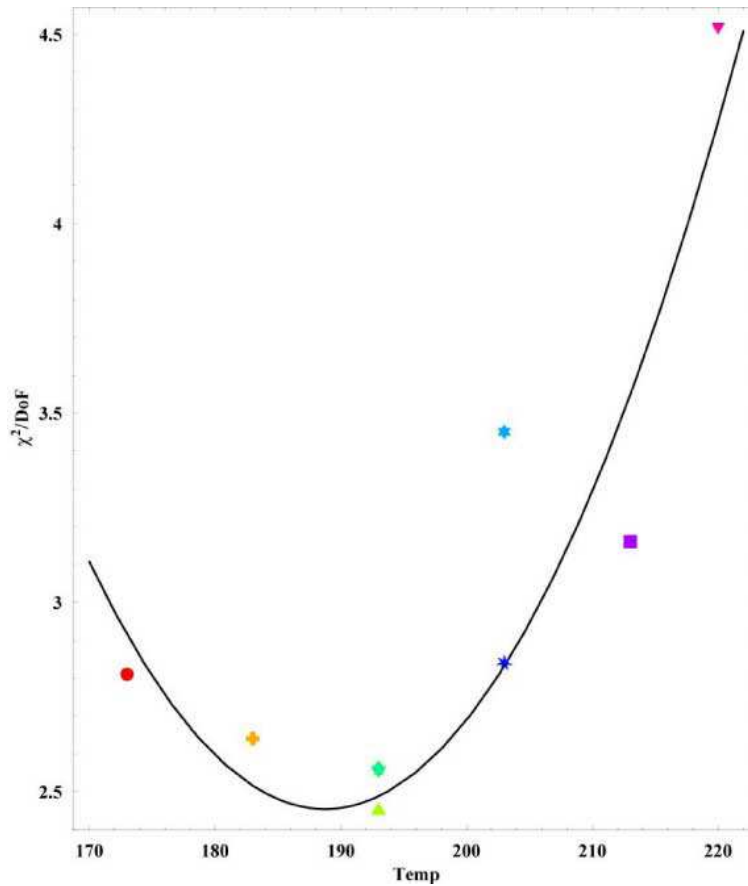}
\caption{\label{temp} (Color online) 
Dependence of chi-squared per degree of freedom on temperature. Symbols correspond sequentially to the fits listed in Table II: circle (red), plus (orange), triangle (green), diamond (chartreuse), 5-star (aqua), 6-star (light blue), 7-star (dark blue), square (violet), and del (magenta).  Black line is a parabolic fit to the points, provided to guide the eye.
}
\end{figure}

Fig. {\ref{temp} shows $\chi^{2}$ per degree of freedom as a function of temperature for the nine fits listed in Table II.  The solid line is a parabolic fit to the points.  We do not assert that there is physics driving the preference of the STAR data for the temperature predicted by lattice gauge calculations, but we believe 
 this  is more than a coincidence.

Fig. {\ref{params} shows the ratio of the parameters of Table II to the F193 parameters, and indicates the range of variations and the correlations of parameters.  Note in particular the correlation between $T$, $a_{WS}$, $w_0$, $\Delta\tau$, and $\eta_{F}$.

Fig. {\ref{fits} shows the predictions of all nine fits as compared with the STAR data.  As can be seen, there are no striking differences.  In order to make the differences more visible, Fig. {\ref{fitsratio} shows the ratio of the predictions of the eight other fits to those of F193.  as c
an be seen the differences in the radius predictions are less than 1\%, while those of the spectrum predictions are as large as 10\%.

However, we note that in the momentum region below 50 MeV/c, where no HBT or spectrum data is available, the different fits make different predictions.  Fig. {\ref{fitslowkt} shows the predictions of the fits of Table II in this region.  We note that the ``wiggles'' arise mainly for the extreme fits, and in particular that these variations are relatively small for the F193 fit that provided the best fit to the higher momentum data.

\begin{figure}
\includegraphics[width=12cm]  {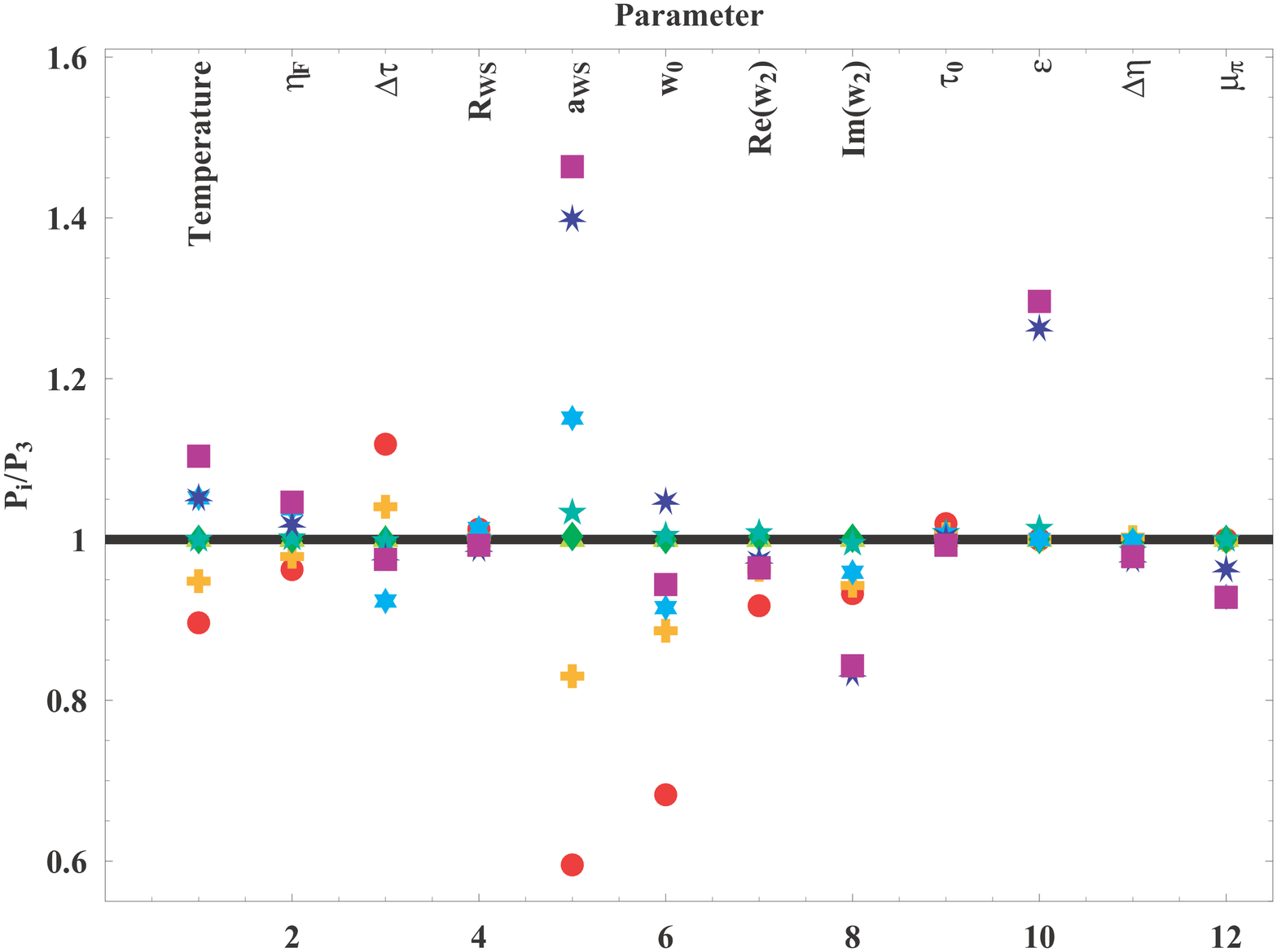}
\caption{\label{params} (Color online) 
Ratio of other fit parameters to those of F193 for the nine fits listed in Table II. Symbols correspond sequentially to the fits listed in Table II: circle (red), plus (orange), triangle (green), diamond (chartreuse), 5-star (aqua), 6-star (light blue), 7-star (dark blue), square (violet), and del (magenta).
}
\end{figure}

\begin{figure}
\includegraphics[width=12cm]  {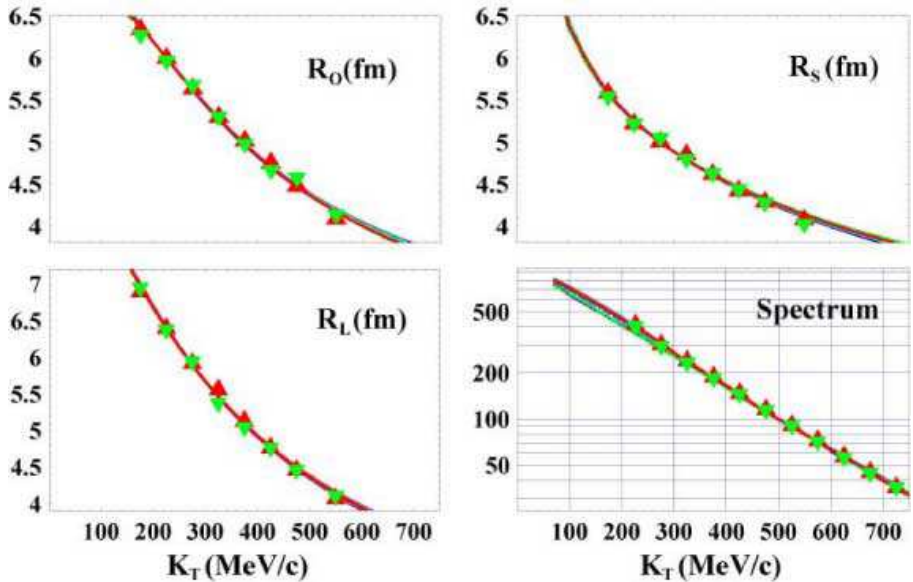}
\caption{\label{fits} (Color online) 
Superposition of nine fits with STAR data, showing that there is relatively little difference between the predictions.  Data points: $\nabla$ (green) $\Rightarrow \pi^{+}$; $\triangle$ (red) $\Rightarrow \pi^{-}$. 
}
\end{figure}

\begin{figure}
\includegraphics[width=12cm]  {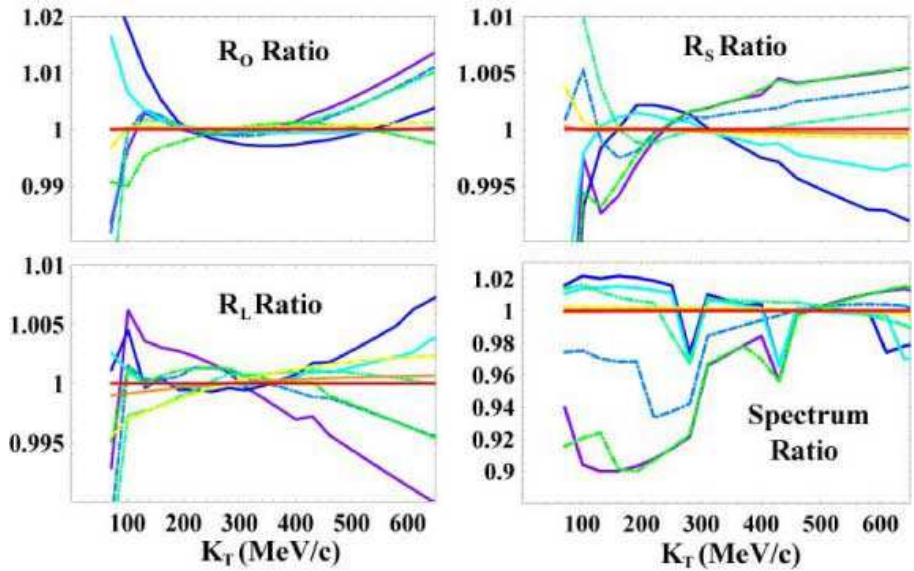}
\caption{\label{fitsratio} (Color online) 
Ratio of other predictions to F193.  Colors correspond sequentially to the fits listed in Table II: red, orange, green, chartreuse, aqua, light blue, dark blue, violet, and magenta.}
\end{figure}

\begin{figure}
\includegraphics[width=13cm]  {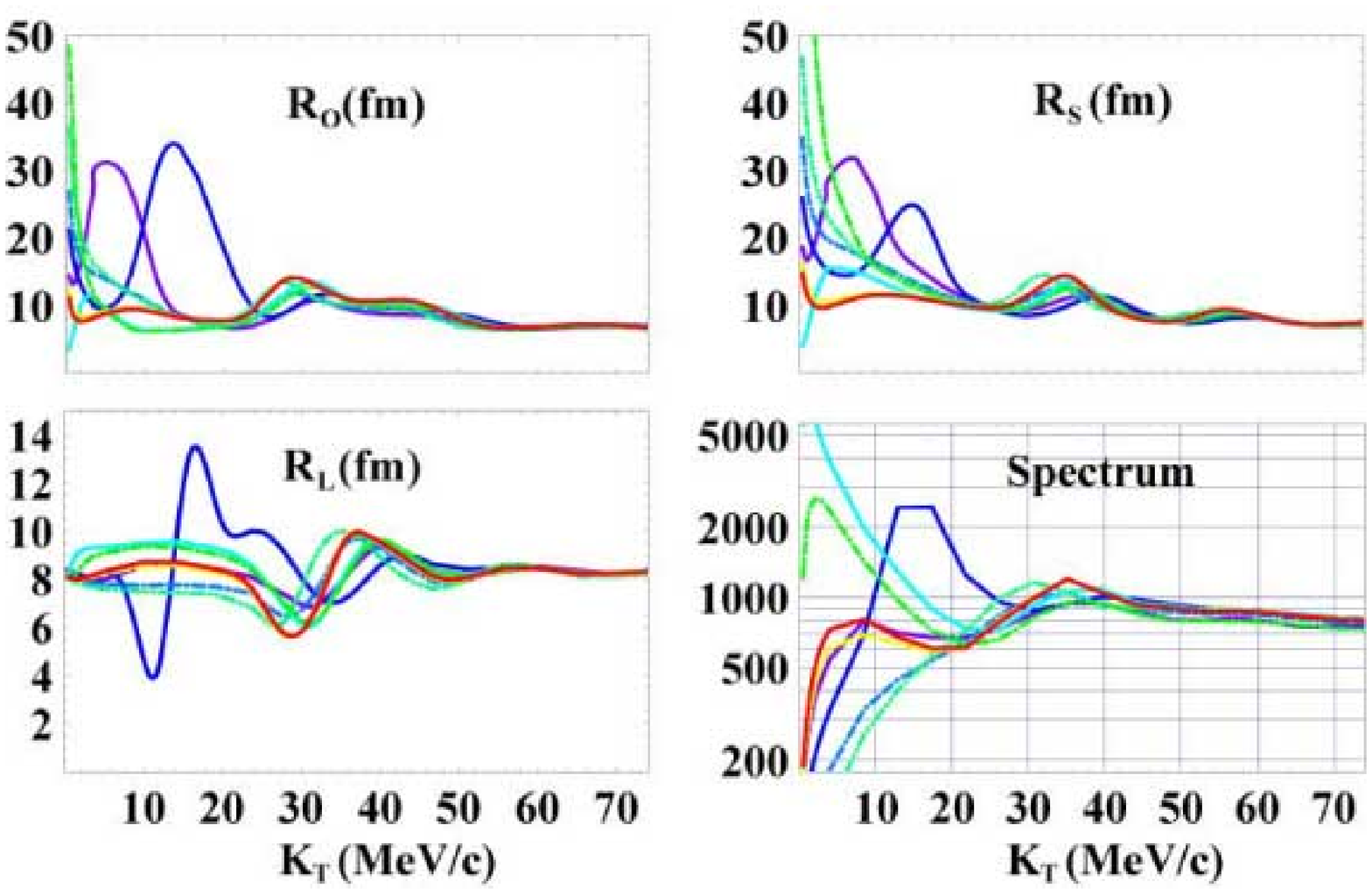}
\caption{\label{fitslowkt} (Color online) 
Low-momentum predictions of the nine fits listed in Table II. Colors correspond sequentially to the fits listed in Table II: red, orange, green, chartreuse, aqua, light blue, dark blue, violet, and magenta.  Note that the low temperature fits (red to aqua) have relatively little structure.
}
\end{figure}

\subsection{ Meaning of optical potential parameters} 
Table I shows that $w_0=0.14$ and that  $w_2$ is 0.85 +0.12 i.  Let us try to understand these values using the
ideas of Sect.~V. First consider the real parts. Comparing \eq{u526} and \eq{uss} shows that
\bea w_0={m_\pi^2\over u_\pi^2} -m_\pi^2(T),\;\; {\rm Re}(w_2)={1-u_\pi^2\over u_\pi^2}\label{ucomp}\eea
which allows us to extract the values:
\bea u_\pi^2=0.54,\;u_\pi=0.74,\; u_\pi m_\pi(T)=0.65\;{\rm fm}^{-1}.\eea
These values are  comparable to the estimates of \cite{Son:2002ci}.

We also compare with the results of Shuryak\cite{Shuryak:1991hb}
for pions propagating in hot dilute matter. His results for $T=200$ MeV (the stated limit of validity
of the calculation) are most relevant for our case.
His Fig.~1 
 shows results for $V=U/(2m_\pi)$ as a function of pion momentum.
The imaginary potential is approximately proportional to $p^2$ for $p\le 400 MeV/c$. For $p=1 $ fm$^{-1}$
Shuryak obtains $Im(U)\approx 0.1$ fm$^{-2}$, which is  close to our value of 0.12 fm$^{-2}$.
 Our real potential
is about seven times larger in magnitude than our imaginary potential, while Shuryak finds that
the strengths are comparable. 
Thus our real potential is much stronger than expected from a dilute gas approximation and  consistent
 (via the
formalism of \cite{Son:2002ci}) with the occurance of a chiral phase transition.

\subsection {Non-central Au+Au }
\label{subs:auau}
Our analysis so far has focused on the central (0-5\%) STAR $\sqrt{S_{NN}}$=200 GeV Au+Au data.  However, the STAR collaboration performed measurements of pion correlations and spectra at $\sqrt{S_{NN}}$=200 GeV Au+Au as
 a function of centrality.

For non-central events, our optical potential 
would depend on the direction of $\bfb$ as well as its magnitude. 
The  simple dependence on $b$ was exploited heavily in previous sections to simplify the calculations, 
so, in principle, non-cental collisions are not a part of our model.  However, we 
can make simplifying assumptions that can allow us to predict the observables for
 non-central collisions.  In particular, we assume 
that a non-central collision resembles a central collision with the same number of participants.  
This assumption allows us to extrapolate our results to systems that do not have perfect
 centrality by using participant scaling. In particular,
we take  the space-time parameters $R_{WS}, a_{WS}$, and $\tau_0$ 
  to scale as the centrality-dependent number of participant particles to the one-third power: 
$N_{\rm part}^{1/3}$. The values of $N_{\rm part}$ are taken from Glauber-model 
calculations\cite{Millerthesis}.
For the Au+Au system, the value of $\Delta\tau$  
is kept at the value of F193, because this is a dynamic quantity 
describing the proper-time duration during which pions are emitted in the collision.  
Table III gives the parameters $R_{WS},\tau_0, a_{WS}$ and $\Delta \tau$ vs. centrality, 
as scaled from the F193 fit of Table I.  
Parameters not listed in the table are the same as those of Table I.

\small
\begin{table*}
  \caption{Au+Au scaled space-time parameters from fit F193}
  \vspace{0.1cm}
  \begin{tabular}{|lllllll|}\hline
   $Centrality$ & $~~~N_{\rm part}$ & $~~~N_{\rm part}^{1/3}$ &~~$R_{WS}$ & $~~a_{WS}$&~~~~~$\tau_{0}$ &~~~~~$\Delta\tau$\\
   & & &~~$(fm)$ &~~$(fm)$&  $~~(fm/c)$ &~~~$(fm/c)$ \\
\hline
{\bf ~~0-5\%}   &~~352.44 &~~~7.064 &~11.773 &~~0.909 &~~~8.601 &~~~~2.382\\
{\bf ~5-10\%}   &~~299.31 &~~~6.689 &~11.149 &~~0.860 &~~~8.145 &~~~~2.382\\
{\bf 10-20\%}   &~~234.55 &~~~6.167 &~10.279 &~~0.793 &~~~7.509 &~~~~2.382\\
{\bf 20-30\%}   &~~166.68 &~~~5.503 &~~9.173 &~~0.708 &~~~6.701 &~~~~2.382\\
{\bf 30-50\%}   &~~~96.07 &~~~4.580 &~~7.634 &~~0.589 &~~~5.577 &~~~~2.382\\
{\bf 50-80\%}   &~~~29.75 &~~~3.099 &~~5.164 &~~0.398 &~~~3.773 &~~~~2.382\\
     \hline
      \end{tabular}
      \end{table*}
\normalsize

\begin{figure}
\includegraphics[width=8 cm]{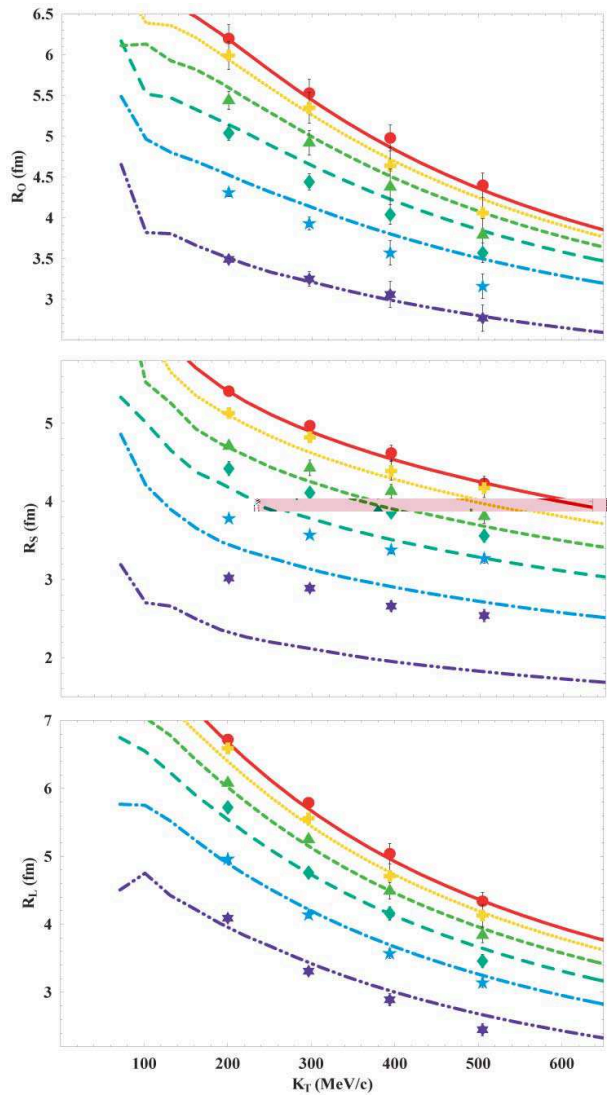}
\caption{\label{Fig510} (Color online) Centrality dependence of HBT radii in Au+Au collisions.  Curves are predictions from participant-scaled space-time parameters; data points are Au+Au at $\sqrt{s_{NN}}=200$ GeV from \cite{STARHBT} .  Symbols and colors: 0-5\% circle (red), 5-10\% plus (yellow), 10-20\% triangle (green), 20-30\% diamond (aqua), 30-50\% 5-star (blue), 50-80\% 6-star (indigo).
}
\end{figure}

In Fig.~\ref{Fig510} we see that the participant-scaled predictions work fairly well for the first few centrality bins, but agreement diminishes for the more peripheral collisions.   We do not expect this simple scaling procedure to be accurate very far from the purely central collision assumption of the model, so this loss of predictive power is to be expected.

In particular, looking for differences between peripheral and central collisions is a standard way to test for the possible existence of QGP effects.  This means that the  strength of the optical potential parameters of our model should change as the collision becomes peripheral. We have kept the potential depth parameters constant with centrality because we have no way to predict their centrality dependence.
Fig.~\ref{Fig510} shows  an excellent reproduction of $ R_L$ with centrality,  a fairly good reproduction of $R_O$, but only a fair description of $ R_S$. We take this to be consistent with
 the disappearance of a chiral phase transition in the more peripheral collisions.


\subsection{Predictions for Cu+Cu  collisions}
\label{subs:cucu}
The STAR collaboration has also measured (but not yet published) 
radii and spectra vs. centrality for Cu+Cu collisions at $\sqrt{S_{NN}}$=200 GeV, so
it is of interest to make predictions for this system. Here  again use
 participant scaling, but 
 have also assumed that the emission 
duration parameter $\Delta\tau$ scales as $A^{1/3}$, 
where $A$ is that atomic mass number of the colliding nuclei.
  Table IV gives the scaled space-time parameters used to predict the HBT radii for the Cu+Cu system. 
  Parameters not present in the table are the same as those of Table I.

\begin{figure}
\includegraphics[width=11cm]{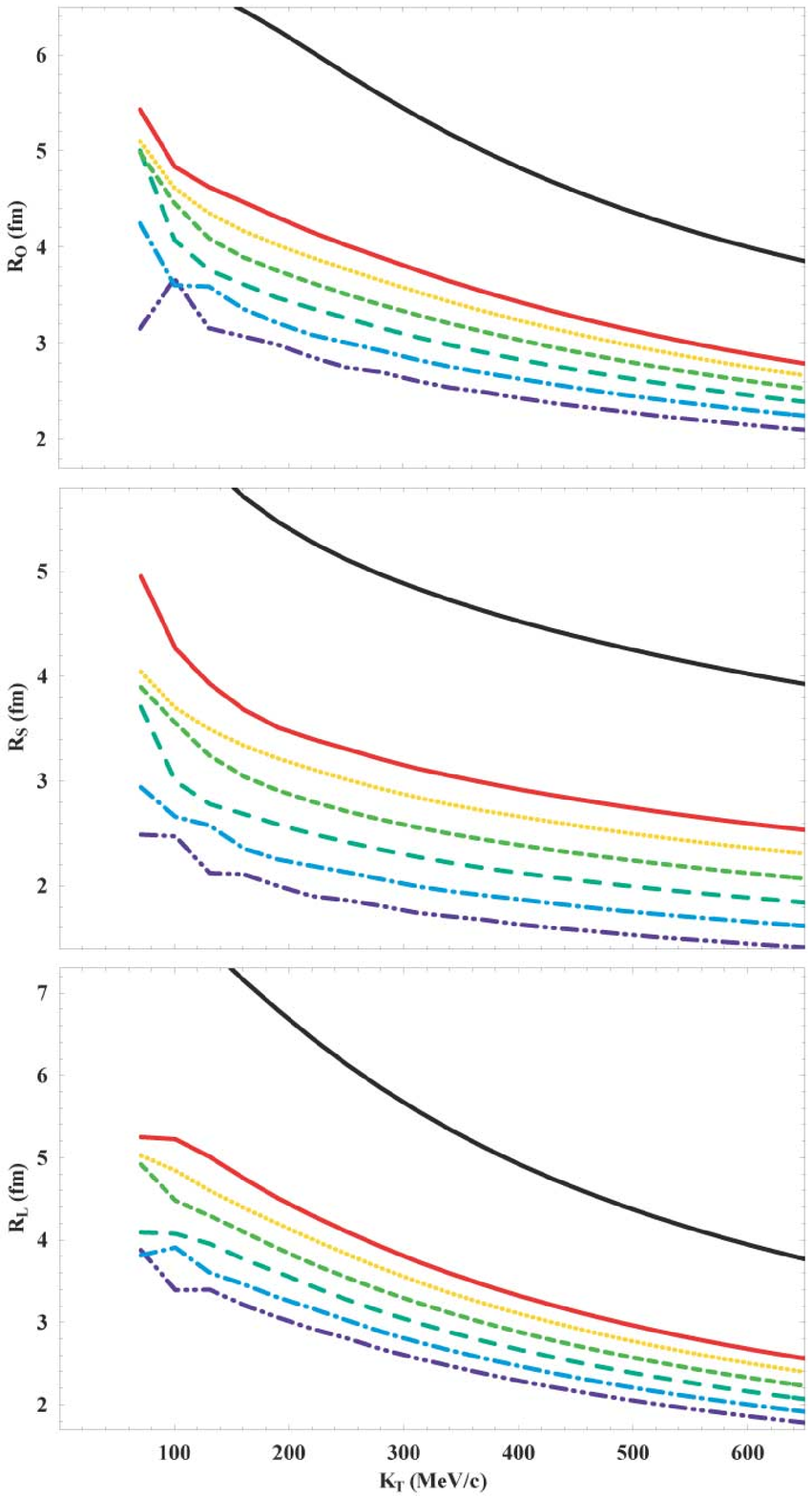} 
\caption{\label{cu} (Color online) HBT Radii: Central Cu+Cu ($\sqrt{s}=200$ GeV) predictions: Au+Au F193 fit -  black solid; Cu+Cu Centrality 0-10\% - red solid, 10-20\% - yellow dot, 20-30\% - green short-dash, 30-40\% - aqua long-dash, 40-50\% - blue dot-dash, 50-60\% - violet dot-dot-dash.} 
\end{figure}

\small
\begin{table*}
  \caption{Cu+Cu scaled space-time parameters from fit F193}
  \vspace{0.1cm}
  \begin{tabular}{|lllllll|}\hline
   $Centrality$ & $~~~N_{\rm part}$ & $~~~N_{\rm part}^{1/3}$ &~~$R_{WS}$ & $~~a_{WS}$&~~~~~$\tau_{0}$ &~~~~~$\Delta\tau$\\
   & & &~~$(fm)$ &~~$(fm)$&  $~~(fm/c)$ &~~~$(fm/c)$ \\
\hline
{\bf ~0-10\%}   &~~98.38 &~~~4.616 &~7.694 &~~0.594 &~~~5.621 &~~~~1.629\\
{\bf 10-20\%}   &~~74.78 &~~~4.210 &~7.022 &~~0.542 &~~~5.130 &~~~~1.629\\
{\bf 20-30\%}   &~~54.37 &~~~3.788 &~6.314 &~~0.487 &~~~4.613 &~~~~1.629\\
{\bf 30-40\%}   &~~38.53 &~~~3.378 &~5.629 &~~0.434 &~~~4.113 &~~~~1.629\\
{\bf 40-50\%}   &~~26.29 &~~~2.974 &~4.956 &~~0.382 &~~~3.621 &~~~~1.629\\
{\bf 50-60\%}   &~~17.63 &~~~2.603 &~4.338 &~~0.335 &~~~3.169 &~~~~1.629\\
     \hline
      \end{tabular}
      \end{table*}
\normalsize

\subsection{Correlation functions and the Gaussian approximation}
\label{subs:cfg}

We apply  the present formalism to obtain correlation function $C$ (with $\lambda=1$),
using the parameters of Table I, for $K_T=158, 316 $ MeV/c. The results are shown in 
Figs.~\ref{corr158316},\ref{FigRa}. We see that the correlation functions $C(K_T,q)$ are 
fairly well represented by
Gaussians, for the relevant range of small values of $q$, 
 with widths that are approximately independent of $K_T$. 
For a typical radius of about
7 fm, and $q=0.15$ fm$^{-1}$, (where data are measured) 
$q^2R^2\approx 1$.  Therefore, using  the  approximation (\ref{ckqquad})
 is not very accurate. On the other hand, 
the correlation functions are close to Gaussian in shape in this region, suggesting that approximation (\ref{ckqgauss1}) is more appropriate.  A detailed look at the ratio of computed correlation functions ($C-1$) 
to its Gaussian fit in Fig.~\ref{FigRa} shows that the Gaussian curves represent the correlation functions
fairly well in the region $0<q<0.22$ fm$^{-1}$ where $C-1$ is large and radii are extracted,
but that the correlation functions are systematically larger than the Gaussian fit for $R_{O,L}$
and smaller for $R_S$  in the ``tail'' region ($q>0.22$ fm).

\begin{figure}
\includegraphics[width=12 cm]{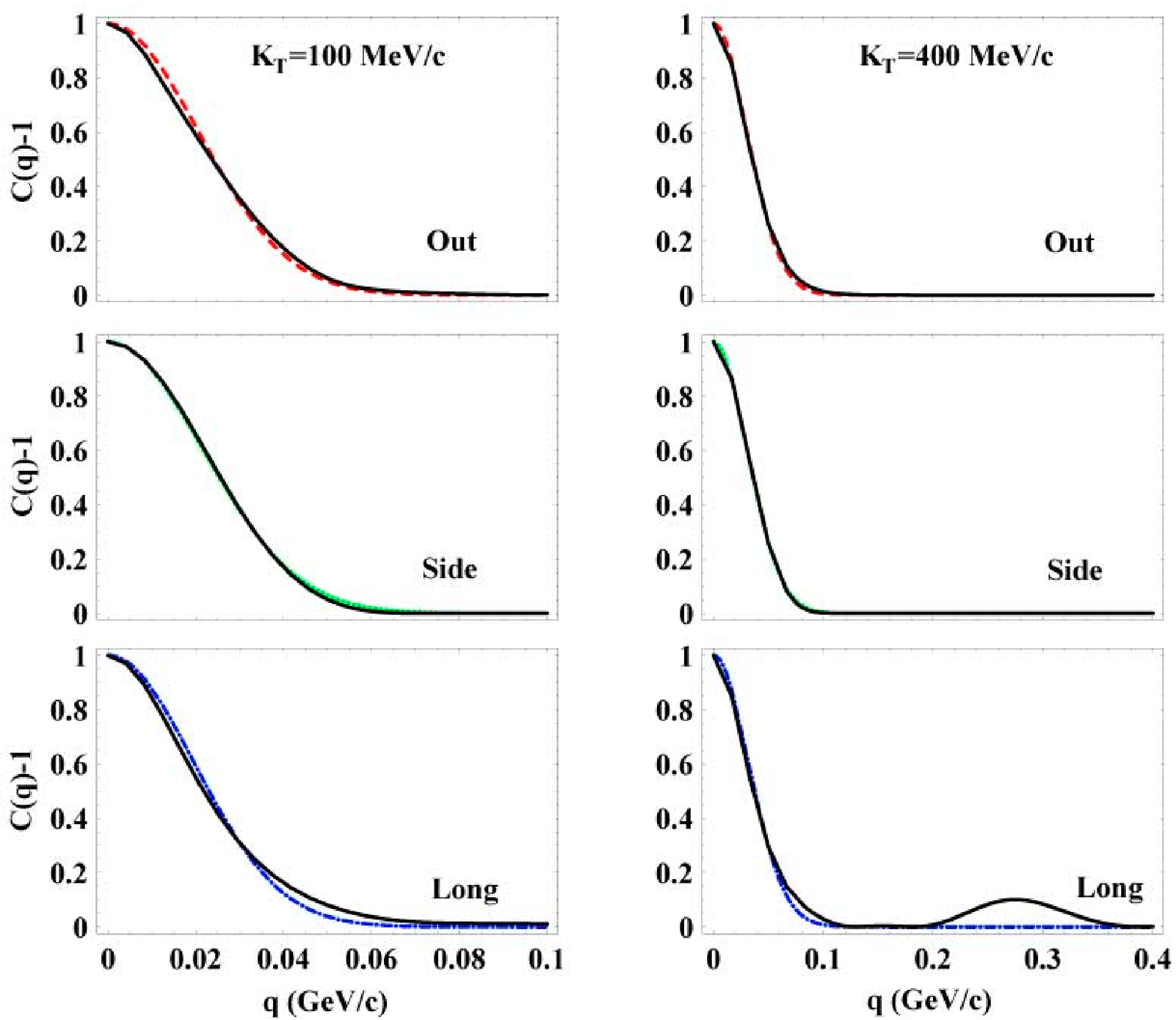}
\caption{\label{corr158316}  (Color online) Central Au+Au ($\sqrt{s}$=200 GeV).
 Correlation functions ($K_T$=158 , 316 MeV/c) 
for the out, side and longitudinal cases.
The solid (black) curves are the full correlation functions.
 The broken curves are Gaussian fits: dash for out, dot for side and dot-dash for longitudinal.}
\end{figure}

\begin{figure}
\includegraphics[width=6 cm]{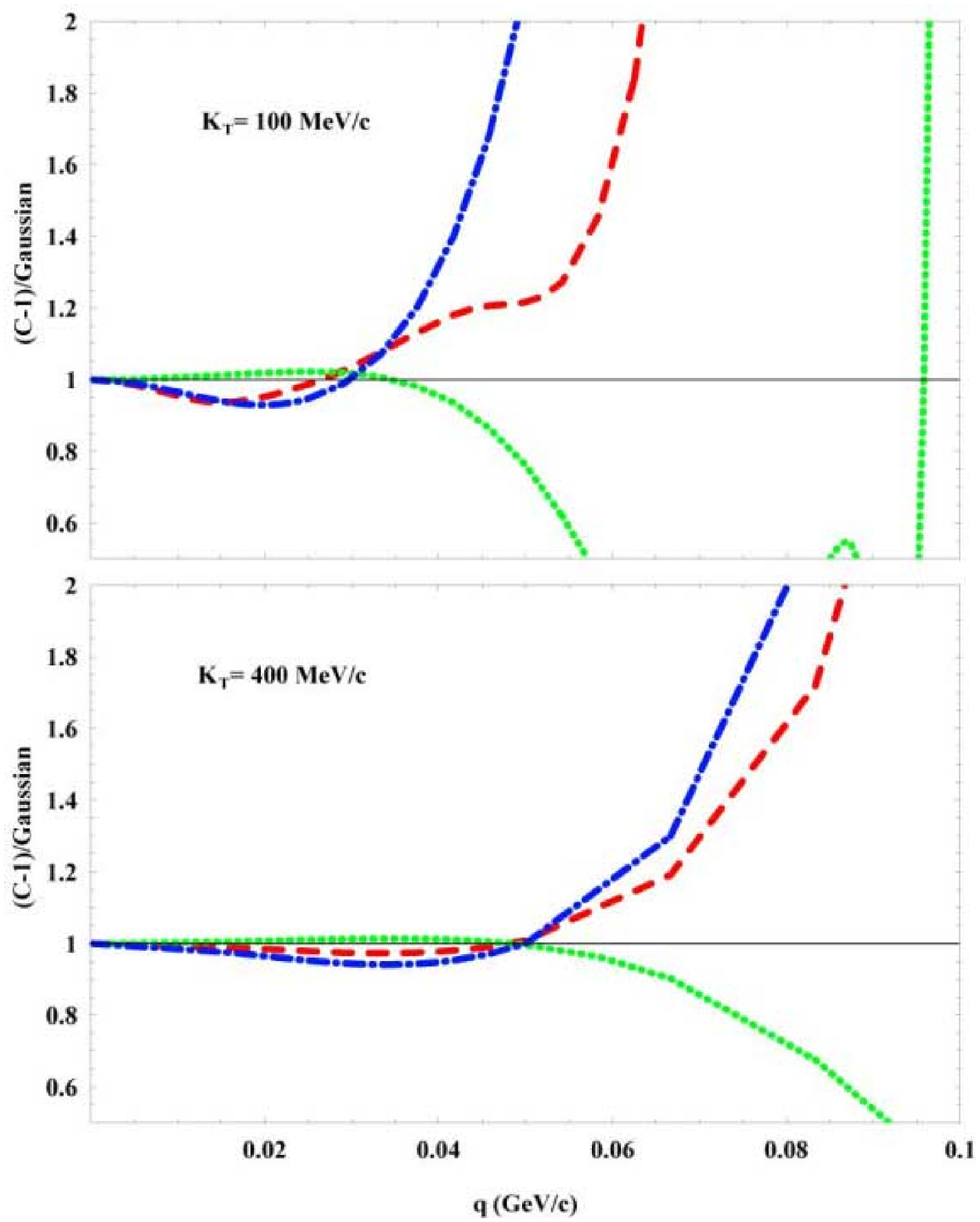} 
\caption{\label{FigRa} (Color Online) Central Au+Au ($\sqrt{s}$=200 GeV). 
Ratios of correlation functions ($K_T$=158, 316 MeV/c) to Gaussian fits 
for the out, side and longitudinal cases.
The dashed (red) curves are obtained with $\bfq$ in the out direction, 
the dotted (green) curves with $\bfq$ in the side direction, and the 
dot-dashed (blue) curves with $\bfq$ in the longitudinal direction. }
\end{figure}

\subsection{Possible Extensions of the DWEF Model}

The DWEF model presented here uses the empirical ``hydrodynamics-inspired'' emission function of Eq.~(\ref{3.1}). However,
we note that the application of distorted waves to an emission function is more generally applicable, and that the formalism we present here can be 
applied to any emission function that has the same symmetry properties as  Eq.~(\ref{3.1}).
One extension of the model would be to calculate the emission function as a multi-dimensional numerical table directly from hydrodynamics and use this with the DWEF model to calculate spectra and radii.
We also note that, while it has not been not implemented here, the DWEF model can allow the temperature and chemical potential to depend on $b$, as is done in the Buda-Lund model \cite{Buda}.  Preliminary investigations of this extension of the model, however, did not lead to an improved description of the data.

\section{The Eikonal Approximation}
\label{sec:eik}

Several previous calculations \cite{Heiselberg:1997vh}--\cite{Wong:2003dh}
of the effects of opacity 
have used the eikonal approximation to
solve the wave equation (\ref{wave}). 
Here we explain the nature of this approximation and discuss its weaknesses and (fewer) strengths
when applied to the current situation.

The basic idea is that if the momentum is large one may say approximately that the
 wave propagate in a given direction
(here the out direction, which is taken as along the $x$ axis). Then one assumes 
 a solution of the
form ${\psi^{(-)}}^*(\bfb)=e^{ipx}\Phi(\bfb)$ that is inserted into
 the wave equation. Taking the Laplacian of the approximate
wave function gives a term proportional to $p^2$ that is canceled, a term proportional to $p$ that
is kept, and another term that is ignored. Then one finds
\bea
{\psi^{(-)}}^*(\bfb=x,y)=e^{ipx}\exp{\left[{-i\over 2p}\int_x^\infty  U(x',y)dx'\right]}\label{eik11}.\eea
for propagation in the $x$ (out) direction. 
The corrections to this solution are of order 
${-1\over 2iK}{1\over U}{\partial U\over \partial x} -{U\over 4K^2}$ times 
the terms  that are kept.
 The first correction can be large in the surface region in which
$U$ varies greatly and
the second term can be large in the interior 
region in which $U$ reaches its full value. 
We are concerned with pions of momentum ranging from 30 to 600 MeV/c, so that 
the eikonal approximation be expected to be poor.
However, the ease of application, and its wide use makes it worthwhile for us to  
 assess the use
of Eq.~(\ref{eik11}). We consider a   purely imaginary potential first and 
then use  a general complex  potential.

\subsection{Strong Absorption at High K -- Purely Imaginary potential}

In the impulse approximation $U=-4\pi f\rho$ where $f$ is the projectile-target
scattering amplitude and $\rho$ is the  density  of scatterers. The optical theorem
relates the imaginary part of $f$ to the total cross section $\sigma$ so that
$Im [U]=-p\sigma \rho$. If we keep only the imaginary potential  $-i/(2p) U= 1/2 \sigma \rho$ and
for a constant density the intensity of the wave falls as $e^{-x\lambda_{\rm mfp}}$ with the mean free path, 
$\lambda_{\rm mfp}={1\over \sigma\rho}$.
More generally 
the  wave function for a purely imaginary optical potential
 is given by
\bea
\psi^{(-)}(\bfp_i,\bfb)=e^{-i\bfp_i\cdot\bfb}e^{-l_i/2\lambda_{\rm mfp}}\label{psiopt}
\eea
where $i=1,2$ for the two wave functions and $l_i(\bfb,\bfK)$ is the 
direct line path length (parallel to the direction of $\bfp_i$
 from the emission point $\bfb$ to the edge of the 
medium. 
 For a purely imaginary optical potential 
$U=-iK\sigma\rho$ and 
 $\lambda_{\rm mfp}$ is the resulting mean free path.

\begin{figure}
\includegraphics[width=4 cm]{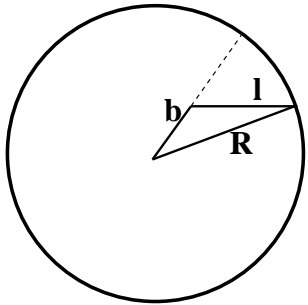}
\caption{\label{eik} 
Eikonal calculation of pion wave function. A pion produced at a position $\bfb$
propagates along a straight line path ${\bf l}$ until it hits the edge of the
medium, denoted by ${\bf R}$. The angle between the
 dashed line and the horizontal line ${\bf l}$ is denoted in the text as $\theta$.
The out case with $\theta_s=0$ is shown.}
\end{figure}

We have, see Fig.~\ref{eik},
\bea
{\bf R}={\bf l_i}+\bfb,\\
R^2=l_i^2+b^2+2 b l_i\cos{(\theta-\theta_s)}.\eea
$\theta$ is angle between $\bfb$ and $x$-axis (direction of $\bfK$)
$\theta_s$ is angle between $\bf p_{1,2}$ and $\bfK$. This was called
$\alpha$ in previous sections. The angle between $\bfb$ and ${\bf l_i}$ is
$\theta-\theta_s$. Solving we find
\bea l=-b\cos(\theta-\theta_s)+\sqrt{(b\cos(\theta-\theta_s))^2+R^2-b^2}.\eea

For the R-out case $\theta_s=0$, and
\bea &&l_i=l=-x+\sqrt{R^2-y^2}\eea 
In this approximation the correlation function minus one is given by
the absolute square of the ratio of integrals: 
${I_{\parallel,\perp}(q)\over I_{\parallel,\perp}(q=0)}.$ In particular,
\bea I_{\parallel}(q)&=&\int d^2b e^{i\bfq\cdot\bfb}e^{-l/\lambda_{\rm mfp}},\quad \bfq\;\|\;{\widehat{\bfx}}\\
&=&{\lambda\over 1+iq\lambda}\int_{-R}^Rdy
\left(e^{iq\sqrt{R^2-y^2}}-e^{-(iq+2/\lambda_{\rm mfp})\sqrt{R^2-y^2}}\right).\label{ipara}\eea
If $\lambda_{\rm mfp}<<R$ we neglect the second term. Since we are interested in radii,
we expand the remaining term in powers of the exponential (keeping 
$\lambda_{\rm mfp}<<R$):
\bea I_{\parallel}(q)
={\lambda_{\rm mfp}\over 1+iq\lambda_{\rm mfp}}\left( 2R-q^2R^2/2(2R-2R/3)+iq\pi R/2\right)\\
C_{\parallel}(q)\approx 2-q^2R^2(2/3-\pi^2/16)\\
R_O^2=R^2(2/3-\pi^2/16)=0.0498R^2; R_O=R/4.48\eea
In the plane wave approximation (Eq.~(\ref{ipara})  with $\lambda_{\rm mfp}\to \infty$) one would 
find $R_O^{PW}=R/\sqrt{8}$. 
The striking result of Heiselberg \& Vischer\cite{Heiselberg:1997vh} indicated that the measured radius
should be 40\%  smaller than the  radius obtained in plane wave approximation.
This result is confirmed for the highest value of $K_T=3~{\rm fm}^{-1}\approx 600$ MeV/c 
by the calculations shown above in Fig.~\ref{rout}. The value of $R_o$ is reduced by
approximately 40\% by the influence of the imaginary optical potential.  Fig.~\ref{imrout}
shows the effects of increasing the
 imaginary potential (by varying $Im [w_2]$ from 0.1 to 0.5 in steps of 0.1). The computed
value of $R_O$ does not change when $Im[w_2]$ is large enough and $K_T$ is high enough.
Small deviations between our results for highest $K_T$ and the  result\cite{Heiselberg:1997vh}
can be attributed to the non-zero value of the diffuseness $a_{WS}$.

\begin{figure}
\includegraphics[width=14 cm]{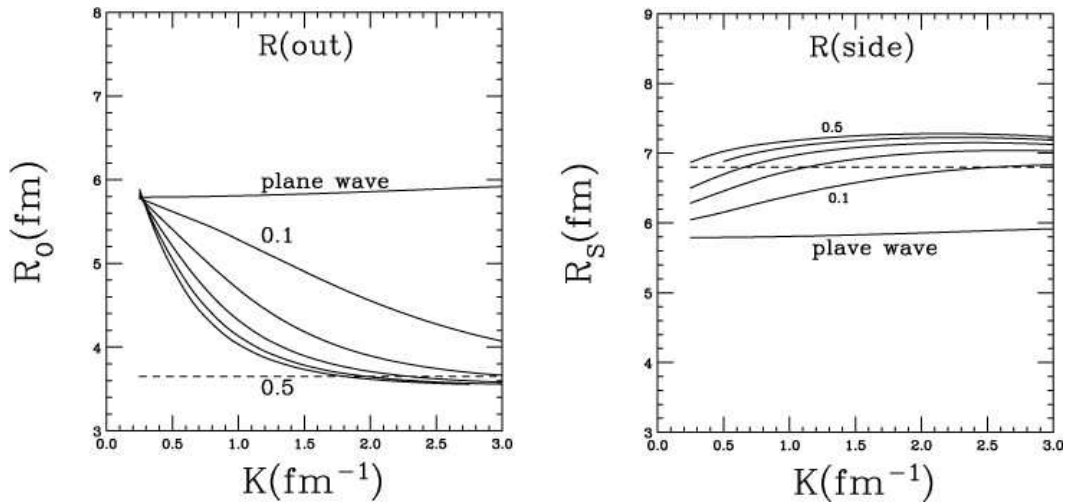} 
\caption{\label{imrout}  Radii $R_o$ and $R_s$ for increasing values of $Im [w_2]$. All other optical potential
parameters and the flow parameters are set to 0. The numbers in the figure refer to the 
range of values. The dashed curve shows the result of Heiselberg \& Vischer. }
\end{figure}

In the model of the present section, 
 the correlation function  and pion intensity would  each be 
proportional to $\lambda_{\rm mfp}$, and a very small value would yield  a very small pionic
spectrum. More generally, the extraction of the chemical potential from the pionic
spectrum 
depends on the mean free path parameter $\lambda_{\rm mfp}$.

Now let us do the R-side case.  Then
\bea \cos\theta_s={K\over \sqrt{K^2+q^2/4}}; 
\sin\theta_s=\pm{q/2\over \sqrt{K^2+q^2/4}}\\
\cos{(\theta-\theta_s)}=\cos\theta(1-{q^2\over8K^2})\mp{q\over 2K}\sin\theta.
\eea
We find that \bea l_i\to-x+\sqrt{R^2-y^2}\pm {\cal O}({q\over K})+ {\cal O}({q^2\over K^2})\approx
-x+\sqrt{R^2-y^2}.\label{moreik}\eea
The terms $\pm {\cal O}({q\over K})$ cancel in computing the term $l_1+l_2$
that enters in computing the present correlation function. Thus the correction is
of order  $1/K^2$. The
eikonal approximation works only if $KR\gg1,$ so the corrections must be 
 presumed  to be small.  
Then 
\bea I_{\perp}(q)
=\int_{-R}^R dy e^{iqy} 
\int_{-\sqrt{R^2-y^2}}^{\sqrt{R^2-y^2}}dx
e^{x/\lambda_{\rm mfp}}e^{-{1\over\lambda_{\rm mfp}}\sqrt{R^2-y^2}}\\
\approx\lambda\int_{-R}^R dy (1-q^2y^2/2)=\lambda_{\rm mfp}(2R-q^2R^3/3)\\
C_\perp(q)=2-q^2R^2/3\\
R_s^2=R^2/3,\eea
Without distortion would be $C^{PW}_\perp(q)=2-q^2R^2/4$, so in this
case the strong absorption increases the radius by a factor of $\sqrt{4/3}.$
This result is qualitatively obeyed in our realistic solutions of the wave equation, see
Fig.~\ref{imrout}.

\subsection{Complex potential with non-vanishing real part}

If there is an attractive real potential the wave function of 
Eq.(\ref{psiopt}) becomes:
\bea
\psi^{(-)}(\bfp_1,\bfb)=e^{-i\bfp_1\cdot\bfb}e^{-l_1(1+i\alpha)/2\lambda_{\rm mfp}}\label{psiopt1}
\eea
with $\alpha$ dimensionless, real and positive. If $\alpha=0$ one obtains the
purely absorptive model of the previous sub-section. Conversely, 
the limit of a purely real potential occurs
when $1/\lambda_{\rm mfp}\to0,\alpha/\lambda_{\rm mfp}\to 1/\lambda_0$.
We also have 
\bea
{{\psi}^*}^{(-)}(\bfp_2,\bfb)=e^{i\bfp_2\cdot\bfb}e^{-l_2(1-i\alpha)/2\lambda_{\rm mfp}}\label{psiopt22}
\eea
In the product  $\psi_{\bfp_1}\psi_{\bfp_2}^*$ enters in computing the correlation
function, so unlike the previous case of pure absorption, a term of the 
form $l_1-l_2$ enters. This difference is of order $q/K$ compared to other
terms, but its influence in computing radii must lead eventually to a term  of 
order $({q\over K})^2$. The validity of the 
eikonal approximation depends on the ability to  disregard such terms. Thus a valid
eikonal approximation means that
 $l_1=l_2$ so 
 the factors of $\alpha$ cancel out in the 
product $\psi^{(-)}(\bfp_1,\bfb){{\psi}^*}^{(-)}(\bfp_2,\bfb)$. Thus, if one assumes the eikonal 
approximation is valid at all values of $K_T$, one would  find erroneously 
that the real potential never has an influence on the 
calculation of  HBT radii. 

Unlike  the effects of the imaginary potential, which  are qualitatively captured by the eikonal
approximation (even if applied wrongly at low values of $K_T$), 
the effects of the real potential are completely lost. Thus the eikonal
approximation can not be used for values of $K_T$ such that the real potential
contributes. As shown in Fig.~\ref{one} the real potential is important for all values of $K_T$
less than  600 MeV/c.  Conversely, for much 
 larger values of $K_T$, for which the eikonal approximation does
accurately reproduce the solution of the wave equation, the real potential will not play 
a role in determining radii.

\section{Oscillations -- a simple square well example}
\label{sec:wigg}

Our numerical results indicate that the radii may have significant oscillations 
for small values of $K_T=K$.   The purpose of this sub-section is to provide a
simple  example that also yields oscillating radii.
 Consider a cylindrical source of radius $R$ of infinite extent in the longitudinal
direction. Suppose this leads to a real,  attractive square 
well potential 
 of radius $R$ that is proportional to the square of the
momentum, as motivated by Eq.~(\ref{u526}).
Then (\ref{wave}) becomes
\bea &&(-\nabla^2 -U_0p^2)\psi=p^2\psi \qquad (b\le R)\\
&&-\nabla^2 \psi=p^2\psi\qquad (b>R), \eea
with $U_0>0$.  This equation is easily solved using the 
partial wave expansion (\ref{psim}). To provide a simple  analytic example
we further specify to the case of
the lowest partial wave, $m=0$. 
 In this case we find
\bea \psi(\bfb)= J_0(p\sqrt{1+U_0}\; b),\qquad(b\le R)\label{m=0}\eea
This shows immediately that the effect of the interaction is to scale each
momenta $p_1,p_2$ that appear in the wave functions   
by a factor $\sqrt{1+U_0}.$ We define \bea \widetilde{p}\equiv p\sqrt{1+U_0}.\eea
 \eq{m=0} 
is a valid approximation to the full
wave function (for $b\le R$) only if $p\sqrt{1+U_0}R\ll1$. However, it is interesting
to also consider larger values of $p$.

To compute radii, recall the correlation function (\ref{cli}).  
For simplicity we take $\eta_f=0$, and 
consider fixed values of $\eta$ and $\tau$, with 
$\Delta \tau=0$. This corresponds to evaluating  $S_0$ at
 its peak and neglecting the
influence of time duration. In this case many factors 
in the numerator and denominator
of Eq.~(\ref{nicecorr}) cancel.  Then 
the correlation function is given by a simple expression
 that provides some insight:
\bea
&&C(K,q)-1=\frac{\phi^2_R(\widetilde{p_1},\tilde{p_2})}{\phi_R(\widetilde{p_1})
\phi_R(\widetilde{p}_2)}\label{ctoy}\\
&&\phi_R(p_i,p_j)\equiv {1\over R^2}\int_0^R bdb J_0(p_ib)J_0(p_jb)=\frac{p_i\,J_0(p_jR)\,J_1(p_iR) - p_j\,
J_0(p_iR)\,J_1(p_jR)}
   {R(p_i^2 - p_j^2)}\label{bark}\\
&&\phi_R(p_i)\equiv{1\over R^2}\int_0^Rbdb J_0^2(p_ib) ={1\over 2}\left(J_0^2(p_iR)+J_1^2(p_iR)\right)
\eea
If we are concerned with the side
radius, inside the well
$p_1=p_2=\sqrt{K^2+q^2/4}$ for the  arguments of $J_0$ that enter.
In that case, $\phi_R(p_1,p_2)\to\phi_R(p_1,p_1)=\phi_R(p_1)$ and the correlation
function takes on the value of 2 and  the side-radius vanishes. 
This is a specific consequence of the approximation of
taking only $m=0$--all directions of $\bfK$ are equivalent, so there is no influence of 
vectors $\bfq$ that are perpendicular to $\bfK$.

The out case is more interesting because the energies and  magnitudes of  
momenta $p_{1,2}=(K\pm  q/2)$ of the two pions are different.
 A non-zero radius is obtained by 
 evaluating
$C(K,q)$ for very small values of $q$ and using \eq{quadform}.
The quantity $R_O(K)/R$ is displayed in Fig.~\ref{figchiral}.

\begin{figure}
\includegraphics[width=14 cm]{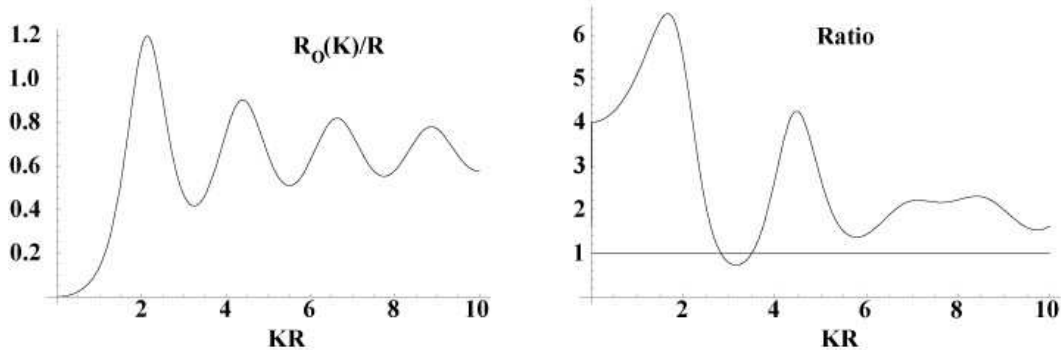} 
\caption{\label{figchiral} $R_{O}(K)/R$ and the ratio of eq (\ref{cratio}). The parameter $U_0=1$.}
\end{figure}
We see that an  oscillatory pattern similar to  our 
realistic results, such as Fig.~\ref{one}
  for small values of $K_T$, emerges from the present calculation. This occurs simply
because Bessel functions oscilate. Indeed the out-radius obtained without the influence of
final state distortions ($U_0=0$) also has oscillations, but with a different frequency.
Therefore 
another quantity of interest is the ratio of radii computed  with $R_O(K)$
and without $R^0_O(K)$ the influence of distortions:
\bea{R_O( K)\over R_O^0(K)}\equiv\;{\rm Ratio}\label{cratio}\eea 
displayed in Fig.~\ref{figchiral}.
The oscillations seen in the right part of 
Fig.~\ref{figchiral} demonstrate the significance of final state 
interactions
that cause enhancements by factors of greater than six  and suppressions by factors of 2!
Furthermore, comparing the left and right hand sides of 
the figure shows the quantity $R^0_O(K)/R$  also   
oscillates. This is caused by the sharp edge
of the square well and because we limit ourselves to $m=0$.

\section{Summary}
\label{sec:sum}
A complete formal treatment of the distorted waves treatment of HBT correlations
is presented here. The need for  incorporating the influence of an optical
potential $U$ and the resulting
distorted wave emission function is explained. The partial wave
formalism necessary to
compute the pionic distorted waves and the resulting emission function is detailed.
Two different  methods with equivalent
 results to evaluate the necessary eight-dimensional integral are described.
Chiral symmetry restricts the form of $U$ \cite{boy,Son:2002ci} at low energy 
and the necessary constraints are 
implemented here and in \cite{Cramer:2004ih}. 
An excellent description of the STAR Au+Au HBT and
spectrum  data is achieved for central collisions and  the use of an 
average area formulation 
leads to a  very good description of these observables for 
non-central collisions.  We also use  
four different versions of geometrical scaling
to  predict the results of central Cu+Cu collisions.
 The ability to calculate the absolute magnitude of 
the spectrum as well as the radii \eq{logform},\eq{quadform}
 (which involve  ratios of functions of emmission functions)
required for the computation of radii is
a principal advantage of our method. The Blast Wave Model is discussed in Sec.~\ref{sec:sss}.

The above results are achieved using a temperature $T_c$ fixed at the recently determined 
critical value of 193 MeV \cite{Katz:2005br}. Such a value could present difficulties
for conventional calculations of the spectra because chemical equilibrium analyses yield
lower temperatures $T_{ch}=174$ MeV \cite{Braun-Munzinger:2001ip}. A large difference between
$T_c$ and $T_{ch}$ implies that the hadrons interact after the deconfinement
transition occurs. This notion is entirely consistent with our treatment of 
pionic distortions which has  as its fundamental assumption that 
pions interact in a hot dense medium before escaping to freedom.

We find that 
the necessary real optical potential is so strongly attractive
that the pion can be said to lose its mass inside the  medium. That chiral symmetry
seems to be restored is the conclusion of our earlier work\cite{Cramer:2004ih}. The RHIC-HBT
puzzle is therefore replaced by the need to investigate this restoration.

Explicit evaluation of wave functions obtained by exact numerical solutions of the wave equation
show some interesting features of the strong interaction and also display differences with
the solutions obtained using the eikonal approximation. A critical discussion of the
eikonal approximation as applied to computing HBT radii shows that its use causes 
the crucial influence
of the real part of the optical potential to be entirely lost.
 The huge importance of the real part of the optical potential is explicitly
illustrated  through a simple 


There are many immediate applications of this formalism. In particular, a
 treatment of HBT data
obtained at all energies is in progress and will be presented elsewhere.

\section*{Acknowledgments}
This work is partially supported by the USDOE grants Nos. DE-FG-02-97ER41014 
and DE-FG-02-97ER41020. GAM thanks LBL, TJNAF and BNL for their
hospitality during the course of this work. We thank W.~Busza,  T.~Cs\"org\H o, J.~Draper, 
M.~Lisa, M.~Luzum, S.~Pratt, J.~Rafelski, S.~Reddy, E.~Shuryak and D. Son for useful
discussions.
\section*{Appendix}
\label{sec:app}
Consider the integral appearing in Sect.~\ref{subs:expeval}:
\bea f(\beta,\Delta\eta)=\int_{-\infty}^\infty d\eta \exp{(-\eta^2/(2\Delta\eta^2))} 
\cosh{(\eta)}\exp{\left({-\beta}\cosh{\eta}\right)},\label{intye}\eea
 that can be approximated \cite{Retiere:2003kf}
using Eq.~(\ref{replace})as
\bea f(\beta,\Delta\eta)=2\exp{({1\over \Delta \eta^2})}K_1(\beta+{1\over \Delta \eta^2}).\label{fap}\eea
Figure~\ref{figftest} shows that the approximation is excellent.

\begin{figure}
\includegraphics[width=12 cm]{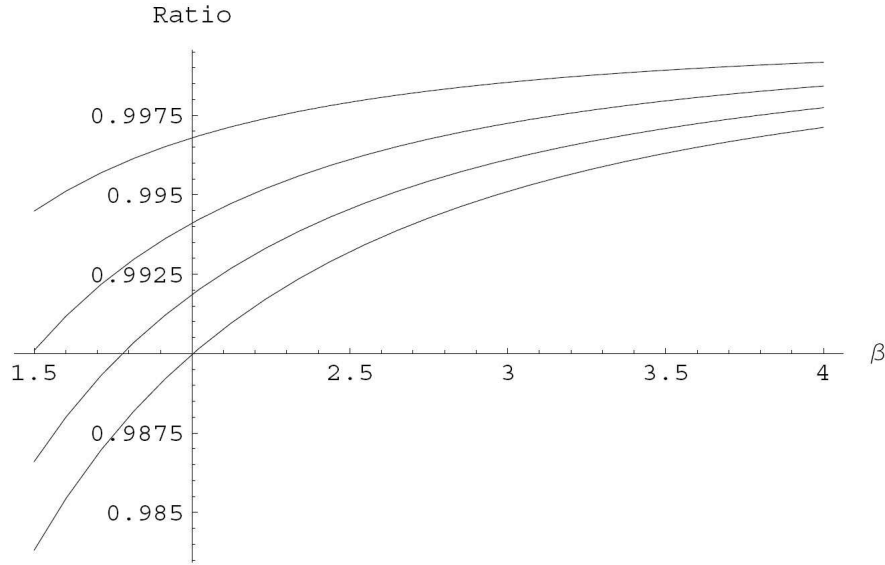} 
\caption{\label{figftest} The ratio of the approximate expression 
 Eq.~(\ref{fap}) to the exact one of   Eq.~(\ref{intye}). The values of $\gamma$ range from
.05 to 0.2, with ratio approaching unity as $\gamma$ approaches 0.}
\end{figure}


\begin{thebibliography}{99}
\bibitem{Pratt:wm}
S.~Pratt,
``Two Particle And Multiparticle Measurements For The Quark - Gluon Plasma,''
 in Hwa, R.C. (ed.): Quark-gluon plasma, vol.2, page 700-748, 1995.

\bibitem{Wiedemann:1999qn}
U.~A.~Wiedemann and U.~W.~Heinz,
Phys.\ Rept.\  {\bf 319}, 145 (1999).
\bibitem{Kolb:2003dz}
P.~F.~Kolb and U.~Heinz,
{\it Quark Gluon Plasma 3}, edited by
 R.C. Hwa and X.-N. Wang, World Scientific, Singapore, 2004)
\bibitem{Lisa:2005dd}
  M.~Lisa, S.~Pratt, R.~Soltz and U.~Wiedemann,
  arXiv:nucl-ex/0505014.
\bibitem{BP-HBT}
S.~Pratt,
Phys.\ Rev.\ Lett.\  {\bf 53}, 1219 (1984).
	G. F. Bertsch {\it et al.}, 
Phys. Rev. {\bf C37}, 1896 (1988).
\bibitem{Rischke:1996em}
D.~H.~Rischke and M.~Gyulassy,
Nucl.\ Phys.\ A {\bf 608}, 479 (1996)

\bibitem{Adler:2001zd}
C.~Adler {\it et al.}  [STAR Collaboration],
Phys.\ Rev.\ Lett.\  {\bf 87}, 082301 (2001);
K.~Adcox {\it et al.}  [PHENIX Collaboration],
Phys.\ Rev.\ Lett.\  {\bf 88}, 192302 (2002);
A.~Enokizono  [PHENIX Collaboration],
Nucl.\ Phys.\ A {\bf 715}, 595 (2003).

\bibitem{Heinz:2002un}
U.~W.~Heinz and P.~F.~Kolb,
hep-ph/0204061.
\bibitem{Gyulassy:2004zy}
  M.~Gyulassy and L.~McLerran,
  Nucl.\ Phys.\ A {\bf 750}, 30 (2005).

\bibitem{Cramer:2004ih}
  J.~G.~Cramer, G.~A.~Miller, J.~M.~S.~Wu and J.~H.~S.~Yoon,
  Phys.\ Rev.\ Lett.\  {\bf 94}, 102302 (2005);  Phys.\ Rev.\ Lett.\ {\bf 95}, 
139901(E) (2005). 

\bibitem{Katz:2005br}
  S.~D.~Katz,
  arXiv:hep-ph/0511166.

\bibitem{Braun-Munzinger:2001ip}
  P.~Braun-Munzinger, D.~Magestro, K.~Redlich and J.~Stachel,
  Phys.\ Lett.\ B {\bf 518}, 41 (2001)
\bibitem{ed73} E.~V.~Shuryak,
  Phys.\ Lett.\ B {\bf 44}, 387 (1973).
\bibitem{GKW79}
M.~Gyulassy, S.~K.~Kauffmann and L.~W.~Wilson,
Phys.\ Rev.\ C {\bf 20}, 2267 (1979).
\bibitem{Kapusta:2005pt}
  J.~I.~Kapusta and Y.~Li,
  arXiv:nucl-th/0503075.
\bibitem{core-halo} T. Cs\"org\H o, B. L\"orstad, and J. Zim\'anyi, Z. Phys. C {\bf 71},
491 (1996).

\bibitem{CL96a}
  T. Cs\"org\H o and B. L\"orstad, Phys. Rev. C {\bf 54}, 1390 (1996)
\bibitem{H96}
  U. Heinz, in: {\it Correlations and Clustering Phenomena in Subatomic
  Physics}, edited by M.N. Harakeh, O. Scholten, and J.H. Koch, NATO ASI 
  Series B, (Plenum, New York, 1997) (Los Alamos eprint archive 
  nucl-th/9609029)
\bibitem{Tomasik:1997eq}
B.~Tomasik and U.~W.~Heinz,
Eur.\ Phys.\ J.\ C {\bf 4}, 327 (1998)
[arXiv:nucl-th/9707001].
\bibitem{Cooper:1974mv}
  F.~Cooper and G.~Frye,
  Phys.\ Rev.\ D {\bf 10}, 186 (1974).

\bibitem{BJD} J.~D.~Bjorken and S.~D.~Drell, {\it Relativistic Quantum Fields} (McGraw-Hill, New York, 1965). 
\bibitem{WSH96}
  U.A. Wiedemann, P. Scotto and U. Heinz, Phys. Rev. C{\bf 53}, 918 (1996) 

\bibitem{Retiere:2003kf}
F.~Retiere and M.~A.~Lisa,
Phys.\ Rev.\ C {\bf 70}, 044907 (2004)     

\bibitem{lsk}
L.~S.~Kisslinger
Phys.\ Rev.\  {\bf98}, 761 (1955)

\bibitem{Buda}
M.~Csanad, T.~Cs\"org\H o, B.~Lorstad and A.~Ster,
J.\ Phys.\ G {\bf 30}, S1079 (2004);
T.~Cs\"org\H o and B.~Lorstad,
Phys.\ Rev.\ C {\bf 54}, 1390 (1996);
M.~Csanad, T.~Cs\"org\H o and B.~Lorstad,
Nucl.\ Phys.\ A {\bf 742}, 80 (2004).
\bibitem{STARHBT}J.~Adams, {\it et al.}  [STAR Collaboration],
Phys.\ Rev.\ C {\bf 71}, 044906 (2005)

\bibitem{Gell-Mann:1968rz}
  M.~Gell-Mann, R.~J.~Oakes and B.~Renner,
  Phys.\ Rev.\  {\bf 175}, 2195 (1968).
\bibitem{Brown:1991kk}
  G.~E.~Brown and M.~Rho,
  Phys.\ Rev.\ Lett.\  {\bf 66}, 2720 (1991).
\bibitem{Koch:1996vt}
  V.~Koch,
{\it Prepared for International Summer School on Correlations and Clustering Phenomena in Subatomic Physics, Dronten, Netherlands, 5-16 Aug 1996}
 V.~Koch,
  Int.\ J.\ Mod.\ Phys.\ E {\bf 6}, 203 (1997).

\bibitem{Son:2002ci}
D.~T.~Son and M.~A.~Stephanov,
Phys.\ Rev.\ D {\bf 66}, 076011 (2002)
D.~T.~Son and M.~A.~Stephanov,
Phys.\ Rev.\ Lett.\  {\bf 88}, 202302 (2002)
\bibitem{boy}
D.~Boyanovsky, H.~J.~de Vega and S.~Y.~Wang,
Nucl.\ Phys.\ A {\bf 741}, 323 (2004).

\bibitem{Baker:1977hp}
G.~A.~Baker, B.~G.~Nickel and D.~I.~Meiron,
Phys.\ Rev.\ B {\bf 17}, 1365 (1978).

\bibitem{sas}C.~Sasaki,
Prog.\ Theor.\ Phys.\ Suppl.\  {\bf 156}, 174 (2004)
\bibitem{STARspec}
J.~Adams, {\it et al.} [STAR Collaboration], Phys. Rev. Lett. {\bf 92}, 112301 (2004).


\bibitem{Akkelin:2003kp}
  S.~V.~Akkelin and Y.~M.~Sinyukov,
  arXiv:nucl-th/0310036.
\bibitem{Back:2004uh}
  B.~B.~Back {\it et al.},
  Phys.\ Rev.\ C {\bf 70}, 051901 (2004).
\bibitem{Shuryak:1991hb}
  E.~V.~Shuryak,
  Nucl.\ Phys.\ A {\bf 533}, 761 (1991).
\bibitem{Millerthesis} M. Miller, Yale University, Ph.D. Thesis 2001
``Measurement of Jets and Jet Quenching at RHIC''
\bibitem{ddd} C.~W.~De Jager, H.~De Vries and C.~De Vries,
  Atom.\ Data Nucl.\ Data Tabl.\  {\bf 36}, 495 (1987).
\bibitem{Chapman:1994yv}
S.~Chapman, P.~Scotto and U.~W.~Heinz,
Phys.\ Rev.\ Lett.\  {\bf 74}, 4400 (1995).


\bibitem{Heiselberg:1997vh}
H.~Heiselberg and A.~P.~Vischer,
Eur.\ Phys.\ J.\ C {\bf 1}, 593 (1998).


\bibitem{Chu:1994de}
  M.~C.~Chu, S.~Gardner, T.~Matsui and R.~Seki,
  Phys.\ Rev.\ C {\bf 50}, 3079 (1994)
  [arXiv:nucl-th/9408005].

\bibitem{Tomasik:1998qt}
B.~Tomasik and U.~W.~Heinz,
arXiv:nucl-th/9805016.
\bibitem{Wong:2003dh}
  C.~Y.~Wong,
  J.\ Phys.\ G {\bf 29}, 2151 (2003).



\end{thebibliography}
\end{document}